\documentclass[a4paper,fleqn]{cas-dc}

\usepackage{aas_macros}
\usepackage[authoryear]{natbib}

\def\tsc#1{\csdef{#1}{\textsc{\lowercase{#1}}\xspace}}
\tsc{WGM}
\tsc{QE}
\tsc{EP}
\tsc{PMS}
\tsc{BEC}
\tsc{DE}

\begin{document}

\let\WriteBookmarks\relax
\def\floatpagepagefraction{1}
\def\textpagefraction{.001}

\shorttitle{One Hundred Years of Venus Polarimetry}

\shortauthors{J Bailey et~al.}

\title [mode = title]{One Hundred Years of Venus Polarimetry: PICSARR Observations of the Phase Curves}                      


%
\author[1]{Jeremy Bailey}[orcid=0000-0002-5726-7000]

\cormark[1]


\ead{j.bailey@unsw.edu.au}

\credit{Writing -- original draft, Writing -- review and editing, Investigation, Methodology, Software, Conceptualization}

\affiliation[1]{organization={School of Physics},
    addressline={University of New South Wales}, 
    city={Sydney},
    postcode={NSW, 2052}, 
    country={Australia}}

\author[2]{Daniel V. Cotton}[orcid=0000-0003-0340-7773]

\credit{Writing -- review and editing, Investigation, Software}

\affiliation[2]{organization={Monterey Institute for Research in Astronomy},
    addressline={200 Eighth Street}, 
    city={Marina},
    state={CA},
    postcode={93933}, 
    country={USA}}

\author[3, 4, 5]{Kimberly Bott}[orcid=0000-0002-4420-0560]

\credit{Writing -- review and editing, Conceptualization, Methodology}

\affiliation[3]{organization={Department of Earth and Planetary Science},
    addressline={University of California}, 
    city={Riverside},
    state={CA},
    postcode={92521}, 
    country={USA}}

\affiliation[4]{organization={NASA Nexus for Exoplanet Science, Virtual Planetary Laboratory Team},
    addressline={Bldg. 3910}, 
    addressline={15th Ave NE, University of Washington},
    city={Seattle},
    state={WA},
    postcode={98195}, 
    country={USA}}

\affiliation[5]{organization={SETI Institute},
    addressline={339 Bernardo Ave, Suite 2000}, 
    city={Mountain View},
    state={CA},
    postcode={94043}, 
    country={USA}}

\author[2, 6]{Ievgeniia Boiko}[orcid=0009-0005-6387-6936]

\credit{Investigation, Writing -- review and editing}

\affiliation[6]{organization={California State University, Long Beach},
    addressline={1250 Bellflower Blvd}, 
    city={Long Beach},
    state={CA},
    postcode={90840}, 
    country={USA}}


\cortext[cor1]{Corresponding author}

\begin{abstract}
We report new high-precision observations of the polarization of light scattered from the atmosphere of Venus, made 100 years after the pioneering studies by Bernard Lyot. The new observations include disk-integrated observations in a range of filters as well as imaging polarimetry. We compare the new results with past observations and models. We have reproduced the 1974 modelling of the Venus polarization by Hansen and Hovenier using modern radiative transfer codes. We show that the new models are in good agreement with the originals, and enable us to calculate the polarization for wavelengths not covered by the original study and to model the polarization distribution across the disk. The new observations are in good agreement with past determinations of the size distribution of the predominant particle mode. They agree with past studies in showing variability of the phase curve between synodic cycles and also polarization variability on short timescales, particularly at higher phase angles (crescent phases). Imaging polarimetry observations show good agreement with models for the redder wavelengths. However, observations in the ultraviolet show very different polarization behavior in the polar regions (within about 30 degrees of the north and south poles). The simplest explanation of this result is that there is a larger Rayleigh scattering component in the polar regions than in the equatorial and mid-latitudes and this could be explained by a lower cloud-top height in agreement with previous spacecraft observations. These ultraviolet polarization observations are inconsistent with horizontally homogeneous atmospheric models. 
\end{abstract}


\begin{highlights}
\item New polarization observations of Venus show that the particle size distribution of the dominant cloud particle mode is unchanged from past observations going back 100 years. 
\item The ultraviolet polarization of Venus in the polar regions is very different to that at lower latitudes and this is most likely due to a lower cloud-top altitude near the poles.
\item The polarization phase curves of Venus are variable on a range of timescales, and continued observations will be needed to cover the full range of variability.
\end{highlights}


\maketitle

\section{Introduction}

The first high quality observations of the polarization of planets were made in the 1920s by Bernard Lyot using telescopes at the Meudon Observatory, Paris. Using a Polariscope and a nulling polarimeter system, Lyot was able to achieve polarimetric precisions of $\sim$0.1\%, which was a remarkable achievement given that the observations were made visually (i.e. by eye). Observations of the polarization phase curves of Venus were made over 1922 to 1924 and the results were reported as ``puzzling and quite different to the polarization of the Moon, Mars and Mercury'' \citep{lyot29}\footnote{An English translation of Lyot's PhD thesis \citep{lyot29} was made by NASA in 1964 as NASA Technical Translation F-187, and is available at: https://archive.org/details/nasa\_techdoc\_19640017154/page/n35/mode/2up}. The polarization varied in a complex way with multiple crossings from positive (perpendicular to the scattering plane) to negative (parallel to the scattering plane).

The results remained puzzling for many years. The single scattering properties of cloud particles could be predicted using Lorenz-Mie theory that was then well understood. However, interpreting the reflection from an atmosphere with multiple scattering required solving a radiative transfer problem including polarization. The methods needed for polarized radiative transfer were not developed until the 1950s and 1960s.

Improved observations of the polarization of Venus were made in the 1950s and 1960s using aperture polarimeters with photomutiplier tubes as detectors \citep{coffeen65,dollfus70}. A key advantage was that the observations could now cover a wide wavelength range from the UV (345 nm) to the near IR (1050 nm).

The development of improved methods for radiative transfer, in particular the doubling method \citep{vanderhulst63}, and its extension to handle polarization \citep{hansen71} allowed progress to be made in understanding the polarization of Venus \citep{hansen71a}. This work culminated in the development of a set of models for the polarization phase curves of Venus at different wavelengths by James Hansen and J.W. Hovenier \citep{hh74}. These models (which we will refer to as HH74) showed that the phase curves could be explained by scattering from optically thick clouds composed of micron-sized liquid droplets with a refractive index consistent with sulfuric acid (75\% H$_2$SO$_4$, 25\% H$_2$O).

Together with observations of the infrared spectrum \citep{young73,pollack74}, the polarization results provided strong evidence for sulfuric acid as the main cloud constituent, a result confirmed by in-situ measurements by the {\it Pioneer Venus} probes \citep{knollenberg80}.

In addition to the ground-based polarization observations described above, the polarization of Venus was observed at four wavelengths by the Orbital Cloud Photopolarimeter (OCPP) instrument on {\it Pioneer Venus} \citep{kawabata80}. Disk-integrated phase curves from these data show similarities to the ground-based data, but the observations also indicate the presence of a haze of sub-micron particles which is particularly important in the polar regions and appears to be variable. Polarization data has also been obtained with the SPICAV-IR instrument on {\it Venus Express} \citep{rossi15}. This data shows the ``glory'' (the negative polarization feature in the IR seen at phase angles of $\sim$15 degrees) and uses it to derive cloud particle sizes in agreement with earlier results.

While the HH74 models \citep{hh74} are reasonably successful in matching the observed polarization phase curves, they are relatively simple models with a single thick cloud layer and a single particle mode. Subsequent studies have shown that the Venus clouds are more complex with multiple particle modes and layers \citep{crisp86}, and these more sophisticated models are needed to explain observations such as night-side infrared spectroscopy that sees deep into the clouds \citep[e.g.][]{meadows96,cotton12,arney14}.

There is now interest in the use of polarization as a means of characterizing the atmospheres of extrasolar planets. Measurements of the disk-integrated polarization of extrasolar terrestrial planets may be feasible with future instruments. The polarization phase curves can provide information that is difficult to obtain in other ways. In particular it may be possible to detect the presence of liquid water through the observation of rainbows \citep{bailey07,karalidi11,karalidi12,sterzik20} or through the detection of glint from oceans \citep{zugger10}. Venus provides a test of some of these methods. It is the only planet in the Solar system with a substantial atmosphere that can be observed through its full range of phase angles from Earth. The rainbow signal, due to liquid sulfuric acid rather than water, is clearly observed.

In this paper we report new observations of the polarization phase curves of Venus, the first such study for more than 50 years. We compare the new observations with past observations and models. We have reproduced the classic HH74 models \citep{hh74} using modern radiative transfer codes, enabling them to be recalculated for wavelengths not investigated in the original work. We also use the observations to obtain polarization images, and compare these with the modelled polarization distribution. We look for changes in the particle properties over time.

\begin{figure*}
    \centering
    \includegraphics[width=17.35cm]{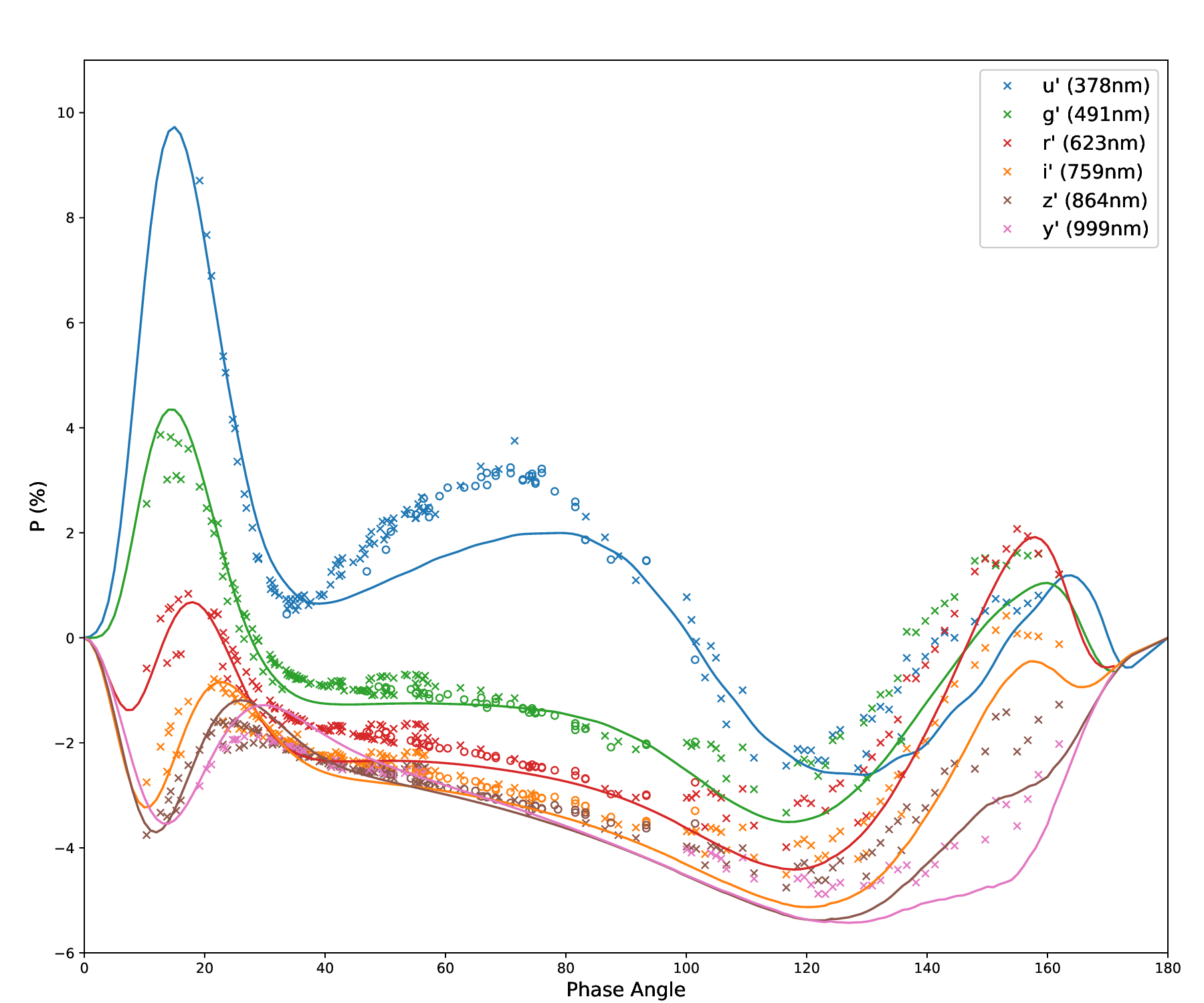}
    \caption{Linear polarization observations of Venus obtained with PICSARR polarimeters over 2021-2024.  Crosses are observations from Pindari Observatory (Sydney, Australia) and circles are from the Weaver Student Observatory (Marina, California). As is conventional for such observations polarization is expressed relative to the scattering plane (the plane containing the Sun, the planet and the observer), such that positive polarization is perpendicular to the plane, and negative polarization is parallel to the plane. The solid lines are based on the HH74 models \citep{hh74} recalculated for our filter wavelengths as described in Section \ref{sec:repro}.}
    \label{fig:picsarr_obs}
\end{figure*}

\section{Observations}

\begin{table*}
\caption{Venus Observation Series}
    \begin{tabular}{lllll}
      Dates   &   Phase angles ($\alpha$) &  Site  &  D/N   &   M/E   \\
\hline
      2021-Aug-13 -- 2021-Dec-17 & 55.0 -- 135.7  &  Pindari   &   Night  &   Evening \\
      2022-Jul-28 -- 2022-Sep-16 & 32.9 -- 13.8   &  Pindari   &   Day    &   Morning \\
      2022-Nov-22 -- 2022-Dec-28 & 10.3 -- 22.2   &  Pindari   &   Day    &   Evening \\
      2023-Jan-11 -- 2023-Mar-30 & 26.9 -- 56.0   &  Pindari   &   Night  &   Evening \\
      2023-Mar-08 -- 2023-Jun-9  & 46.9 -- 93.4   &  WSO       &   Night  &   Evening \\
      2023-Jun-18 -- 2023-Aug-6  & 100.1 -- 161.9 &  Pindari   &   Night  &   Evening \\
      2024-Jul-23 -- 2024-Oct-30 & 19.1 -- 56.6   &  Pindari   &   Night  &   Evening \\
\hline      
    \end{tabular}
\label{tab:series}    
\end{table*}

\begin{table}
    \centering
    \caption{Filter Properties.}
    \begin{tabular}{l|l|l}
    \hline
    Filter & $\lambda$ range (nm)$^a$ & $\lambda_{\rm eff}$$^b$ \\
\hline     
    u$^\prime$ & 320 -- 385  &  378$^c$  \\
    g$^\prime$ & 401 -- 550  &  491  \\
    r$^\prime$ & 562 -- 695  &  623  \\
    i$^\prime$ & 695 -- 844  & 759  \\
    z$_s^\prime$ & 826 -- 920  &  864 \\
    z$^\prime$ & 825 -- $\sim$1100$^d$  &  889  \\
    $y^\prime$  & 950 -- 1058 &  999 \\
    \hline
    \end{tabular}
    \label{tab:filters}
\begin{flushleft}
Notes: \\
$^a$ Measured at 50\% transmission. \\
$^b$ Typical value for Venus observations. \\
$^c$ Short wavelengths are cut off by the detector response. \\
$^d$ Used at WSO only. The long wavelength cut off is set by the detector, not the filter. \\
\end{flushleft}    
\end{table}

Venus is bright and therefore its disk-integrated polarization can easily be measured with small telescopes fitted with suitable instrumentation. The PICSARR (Polarimeter using Imaging CMOS Sensor And Rotating Retarder) instruments used here can be mounted on small telescopes and provide high precision \citep[$\sim10$ parts-per-million in fractional polarization,][]{bailey23}.

For these observations, we used two telescopes at two observing sites. A 20~cm aperture classical Cassegrain telescope was used at Pindari Observatory in Sydney, Australia. A 35~cm Schmidt-Cassegrain telescope (Celestron 14-inch) was used at the Weaver Student Observatory (WSO, \citealp{babcock08}) of the Monterey Institute for Research in Astronomy, located in Marina, California. Most of the observations used the PICSARR instruments as described by \cite{bailey23}. Observations in 2024 used an upgraded instrument in which a direct-drive hollow-shafted motor was used for waveplate rotation, and the ASI 290MM camera was replaced by the ASI 462MM which has higher sensitivity and lower noise.

The observations were taken over 2021 to 2024 and cover about 3 years and two full synodic cycles (a synodic cycle is 584 days). The observations can be broken down into several distinct series as listed in Table \ref{tab:series}. Most of the observations were ``Night'' observations, meaning that the Sun was below the horizon. Many of these observations are actually in twilight, but the sky brightness is sufficiently low that it is very much lower than the surface brightness of Venus. PICSARR's double image design largely eliminates sky background polarization due to the overlap of the double images, and any residual background is subtracted in the aperture extraction process \citep{bailey23}. 

To observe Venus at small phase angles it is necessary to observe in daylight with Venus close to the Sun (``Day'' observations in Table \ref{tab:series}). The sky background is then larger and comparable to the surface brightness of Venus. Under these conditions, scattered sunlight within the telescope and instrument can exacerbate the background problem. To minimize these problems, we fitted the telescope with additional baffles and a sunshade which reduced the aperture of the telescope from 20~cm to 9.5~cm, and were arranged to prevent direct sunlight from reaching the telescope mirror. The sunshade was constructed in modular 3D printed plastic sections, allowing it to be extended in length as Venus came closer to the Sun. In this way, we were able to obtain observations down to an elongation of 7.5$^\circ$ from the Sun.

The observations were made using Sloan Digital Sky Survey (SDSS) filters with properties listed in Table \ref{tab:filters}. The Pindari observations used the Astrodon Generation 2 SDSS filter set. The WSO observations used filters from Chroma \footnote{www.chroma.com}. The $u^\prime$, $g^\prime$, $r^\prime$ and $i^\prime$ filters from both manufacturers are very similar. The $z_s^\prime$ filter used for the Pindari observations has a long-wavelength cut-off at 920 nm. The Chroma $z^\prime$ filter used for WSO observations is open-ended at long wavelengths and uses the detector to set the long-wavelength limit, resulting in a longer effective wavelength than the $z_s^\prime$ filter. The observations from 2023 June onwards include the $y^\prime$ filter at a wavelength of 999 nm. A small number of observations were made in a Bessell B filter (effective wavelength $\sim$ 460 nm). Since they cover only a limited phase range they were not used in the analysis except for that of section \ref{sec:seas_var}, but are included in the table of observations.

The effective wavelength of each observation was determined using a bandpass model \citep{bailey20a} that takes account of the source spectrum, the instrumental response and the Earth atmosphere transmission. For the Venus spectrum, we used a solar spectrum modified by the Venus albedo \citep{barker75}. Table \ref{tab:filters} lists the average effective wavelength for each filter. The effective wavelengths of each individual observation are listed in the table of observations and may differ somewhat from these averages as a result of the airmass of observation and instrumental changes.

The polarization position angle of the observations was calibrated using observations of highly polarized standard stars. The calibration includes empirical corrections for the wavelength dependence of the fast-axis orientation of the superachromatic half-wave plates used in PICSARR and uses data on standard stars from \citet{cotton24}. For the disk integrated observations shown in Fig. \ref{fig:picsarr_obs} the polarization was then rotated to give that relative to the scattering plane (the plane containing the Sun, the planet and the observer) using data from the JPL Horizons ephermeris system \citep{giorgini15}. For imaging data a similar procedure is used, but the final rotation is done by minimizing the integrated signal in the U Stokes parameter. 

\section{Discussion}

\subsection{The Hansen \& Hovenier (HH74) Models}

\label{sec:models}

The models of HH74 \citep{hh74} represent the Venus clouds as a single optically thick layer. However, within this layer, the models include the effects of three different components.

Firstly, there are spherical particles with a gamma size distribution \citep{hansen74b}, an effective radius $r_{\rm eff}$ = 1.05 $\mu$m and effective variance $v_{\rm eff}$ = 0.07. The refractive index varies from 1.45 to 1.43 depending on wavelength and is consistent with that of sulfuric acid (75\% H$_2$SO$_4$, 25\% H$_2$O). These particles are now referred to as ``Mode-2'' cloud particles \citep[e.g.][]{crisp86} and are the predominant particle type near the cloud tops. 

The second component is a population of Rayleigh scatterers mixed with the cloud particles and with an optical depth set to a fraction f$_R$ = 0.045 of the cloud optical depth at a wavelength of 365 nm. HH74 interpret this component as Rayleigh scattering due to the atmospheric gas (primarily CO$_2$) and use the measured ratio f$_R$ to determine that the atmospheric pressure at the cloud tops (defined by an optical depth $\tau$ = 1) is 50 mbar.
Based on the Venus International Reference Atmosphere \citep[VIRA,][]{seiff85} a pressure of 50 mbar corresponds to a height of 68 km at most latitudes. This is reasonably consistent with more recent determinations of the cloud top height \cite[e.g.][]{lee12,federova16,sato20}.

The third component in the HH74 models is less obvious, but was introduced into the models by adjusting the single scattering albedo $\varpi_0$ of the clouds to a wavelength dependent value which gave a spherical albedo for Venus in agreement with that observed. The cloud particles are modelled with a zero imaginary refractive index, which should result in pure scattering particles with no absorption. However, the adjustment of the single scattering albedo, as described by HH74, makes the particles absorbing, particularly at short wavelengths. In effect, this incorporates into the model what is now referred to as the ``unknown UV absorber'' \citep{pollack80}. The UV absorber is treated as a pure absorber with no direct polarizing effect, but it alters the percentage polarization through its effect on the total flux.

\begin{figure}
    \centering
    \includegraphics[width=\columnwidth]{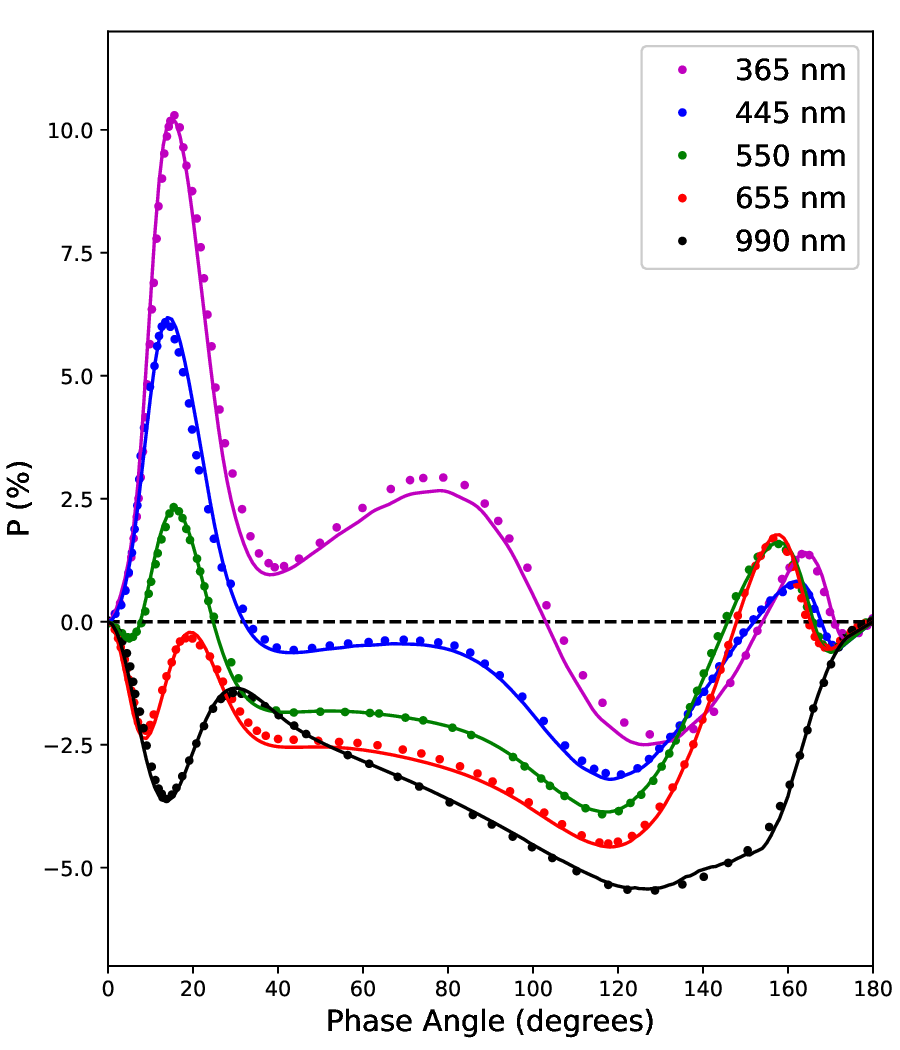}
    \caption{Comparison of models of the Venus disk-integrated polarization from HH74 \citep{hh74} and using our version of the same single layer model using VSTAR/VLIDORT \citep{bailey18}. Circles are models from HH74, digitized from their published figures. Solid lines are equivalent models recalculated using VSTAR/VLIDORT as described in Section \ref{sec:repro} and using the parameters listed in Table \ref{tab:hh_models}.}
    \label{fig:model_comp}
\end{figure}

\begin{table}
    \centering
    \caption{Parameters of models used to reproduce those in HH74 figures.}
    \begin{tabular}{l|l|l|l}
    \hline
    Wavelength (nm) & Figure in HH74 & $n_r$ & $\varpi_0$ \\
\hline     
    365  & Fig. 9 & 1.45$^a$ & 0.98427  \\
    445  & Fig. 11 & 1.44 & 0.99500  \\
    550  & Fig. 4 & 1.44 & 0.99897 \\
    655  & Fig. 12 & 1.44 & 0.99930 \\
    990  & Fig. 7  &  1.43 & 0.99941 \\
    \hline
    \end{tabular}
    \label{tab:hh_models}
\begin{flushleft}
Notes: \\
$^a$ The value of 1.45 is that given in the figure. Elsewhere in the HH74 text the value for this wavelength is given as 1.46. \\
\end{flushleft}    
\end{table}

\begin{table}
    \centering
    \caption{Parameters of models used to match our filter wavelengths.}
    \begin{tabular}{l|l|l|l}
    \hline
    Wavelength (nm) & Filter & $n_r$ & $\varpi_0$ \\
\hline     
    378  & $u^\prime$ & 1.4495 & 0.9870  \\
    491  & $g^\prime$ & 1.4413 & 0.9974  \\
    623  & $r^\prime$ & 1.4359 & 0.9994 \\
    759  & $i^\prime$ & 1.4322 & 0.9994 \\
    864  & $z_s^\prime$ & 1.4302 & 0.9994 \\
    999  & $y^\prime$ & 1.4283 & 0.9994 \\
    \hline
    \end{tabular}
    \label{tab:fil_models}
\end{table}

\begin{figure*}
    \centering
    \includegraphics[width=17.35cm]{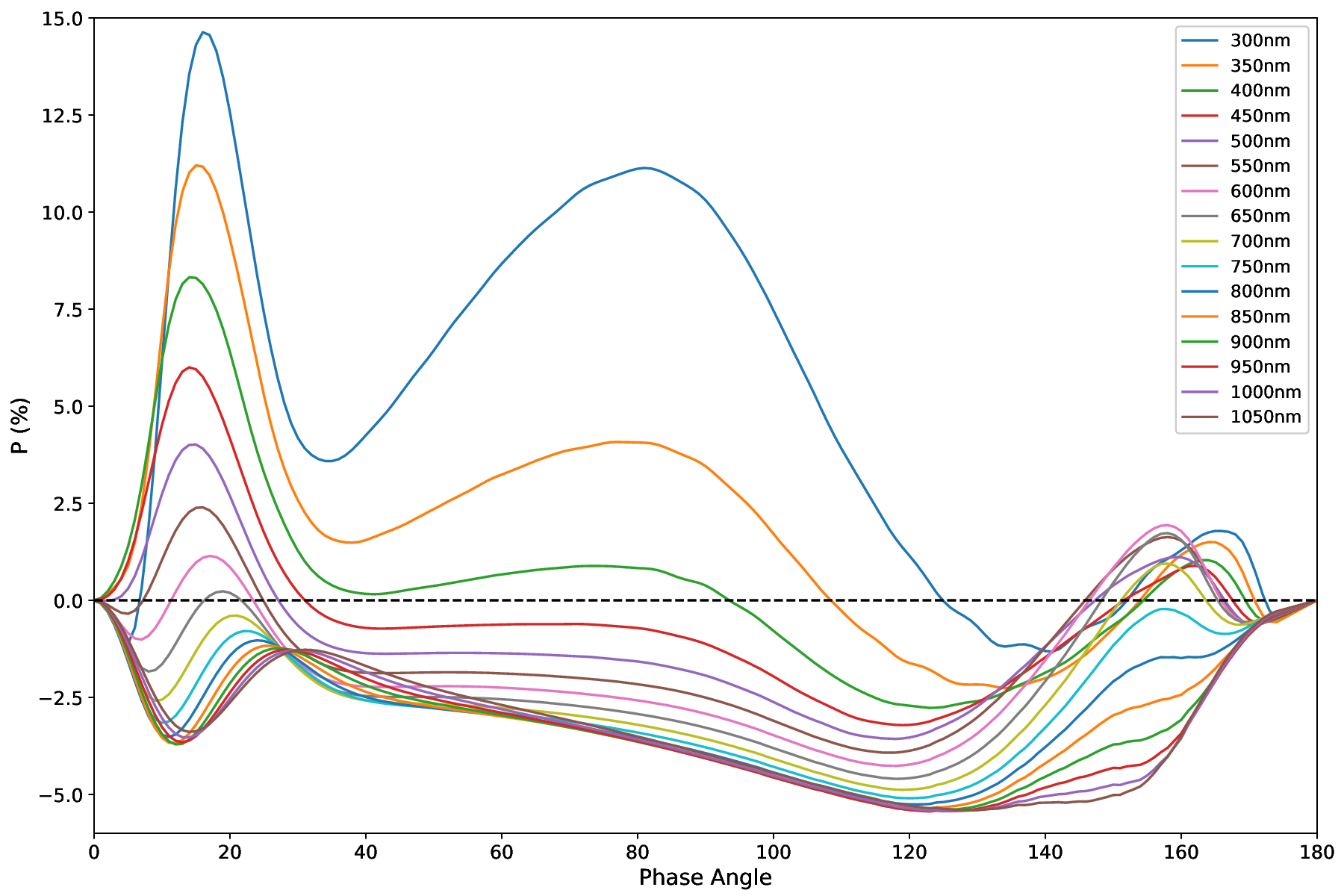}
    \caption{Polarization models calculated with VSTAR/VLIDORT for wavelengths from 300~nm to 1050~nm as described in Section \ref{sec:repro}. The strong peak at the left (phase angle $\sim$ 15 to 20 degrees) is the primary rainbow from the cloud particles. The peak at $\sim$80 degrees in UV wavelengths is mostly due to Rayleigh scattering.}
    \label{fig:mod_wavelengths}
\end{figure*}

\subsection{Reproduction of the HH74 Modelling}
\label{sec:repro}

We have reproduced the HH74 modelling \citep{bailey18} using the VSTAR planetary atmosphere code \citep{bailey12} combined with the VLIDORT polarized radiative transfer code \citep{spurr06}. Our modelling aims to reproduce as closely as possible the original HH74 models. Full details of the methods used and results for three wavelengths 
are given by \cite{bailey18}. Here we present a more complete set of models including models recalculated for our filter wavelengths which are different to the wavelengths presented in the original HH74 work.

In order to reproduce the models of HH74, there are two wavelength-dependent parameters that are needed. These are the real component of the refractive index $n_r$, and the single scattering albedo $\varpi_0$. The exact values of single scattering albedo used by HH74 are listed in their paper only for three wavelengths (365, 550 and 990 nm). While they describe how the single scattering albedo is related to the spherical albedo of Venus, the values they used for other wavelengths are not stated. For other parameters we adopt the HH74 final values of $r_{\rm eff}$ = 1.05 $\mu$m, $v_{\rm eff}$ = 0.07 and f$_R$ = 0.045.

In our first set of models we aim to reproduce the results presented in the figures of HH74 for five wavelengths enabling a direct comparison of our modelling with theirs. The parameters of these models are listed in Table \ref{tab:hh_models}. The single scattering albedos for 365, 550 and 990~nm are those given by HH74. For the two other wavelengths we have adjusted the single scattering albedo to provide the best match to the HH74 modelling. In particular the height of the rainbow feature (phase angle $\sim$ 15 to 20 degrees) in the 445 nm model is sensitive to the single scattering albedo.

In Fig. \ref{fig:model_comp} we show a direct comparison of the polarization of Venus as modelled by HH74 with our results from VSTAR/VLIDORT at the same wavelengths. The HH74 data were digitized from the figures in their paper and are plotted as circles. The corresponding VSTAR/VLIDORT modelling is plotted as solid lines. In general there is good agreement between the two sets of modelling. There are some small differences, largest in the 365~nm models. There is no obvious systematic trend with wavelength, and it may be that the differences are due to inaccuracies in the original plots or the digitization process. It is also possible that small differences are the result of approximations involved in the different radiative transfer methods (the doubling method by HH74, and the discrete-ordinate method in VLIDORT) or the different approaches to integration over the disk.

In order to calculate models at wavelengths different from those included in the HH74 analysis, we have adopted a set of smoothly variable values for $n_r$ and $\varpi_0$, based on the trends in the values shown in Table \ref{tab:hh_models}, that cover the wavelength range from 300~nm to 1050~nm. Fig. \ref{fig:mod_wavelengths} shows the polarization phase curves for these models for a set of wavelengths over this range with 50~nm spacings. We have also calculated models to match the wavelengths of the PICSARR filters used for our observations. The parameters of these models are listed in Table \ref{tab:fil_models} and it is this set of models that are plotted on Fig. \ref{fig:picsarr_obs}.

\begin{figure*}
    \centering
    \includegraphics[width=17.35cm]{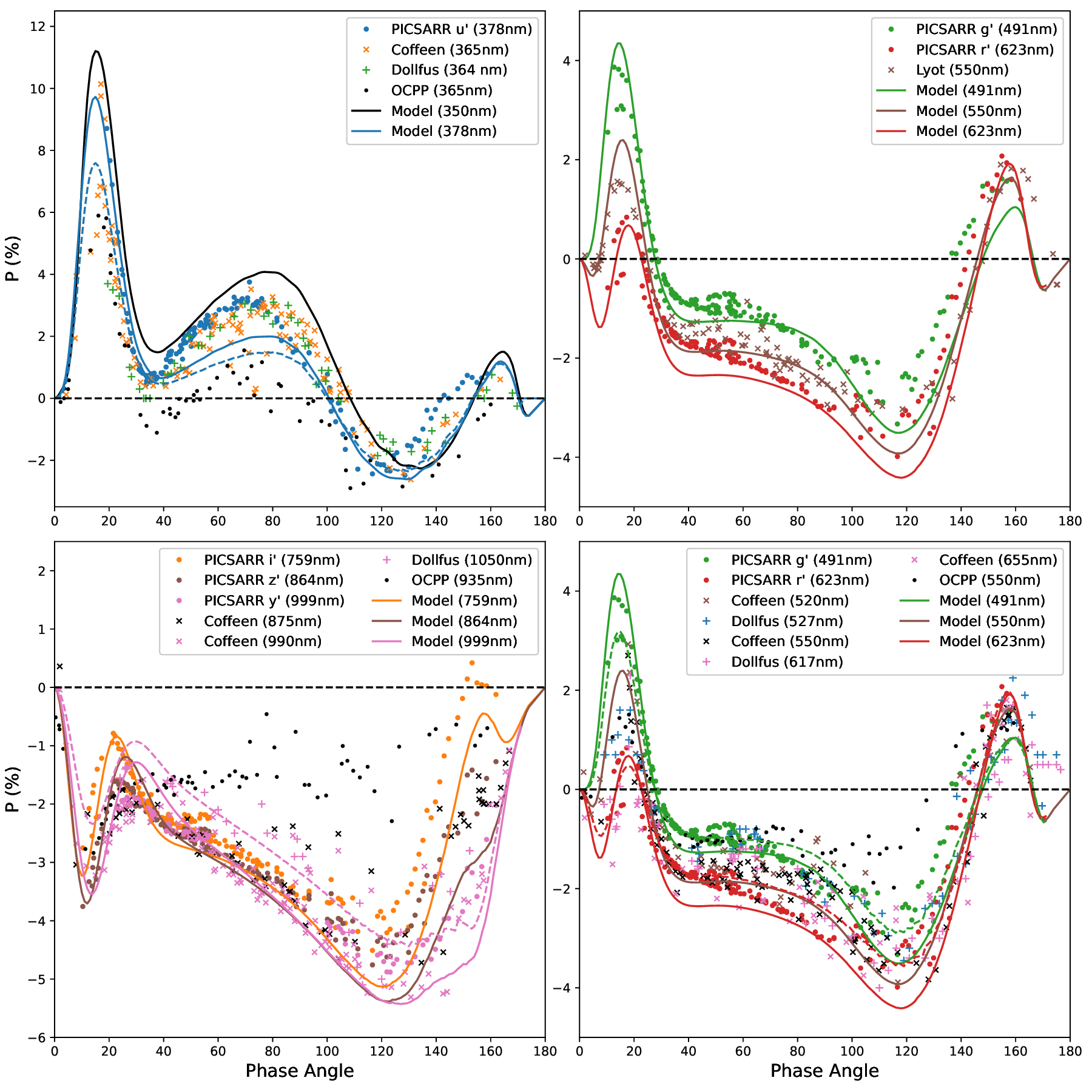}
    \caption{Comparison of new PICSARR observations with selected past observations and models in different wavelength regions. Solid lines are disk-integrated polarization models selected from those used in Figs \ref{fig:picsarr_obs} and \ref{fig:mod_wavelengths}. For 3 wavelengths (378nm, 491nm and 999nm) we also plot as dashed lines the modelled polarization of the center of the image.}
    \label{fig:wavelengths}
\end{figure*}

\begin{figure}
    \centering
    \includegraphics[width=\columnwidth]{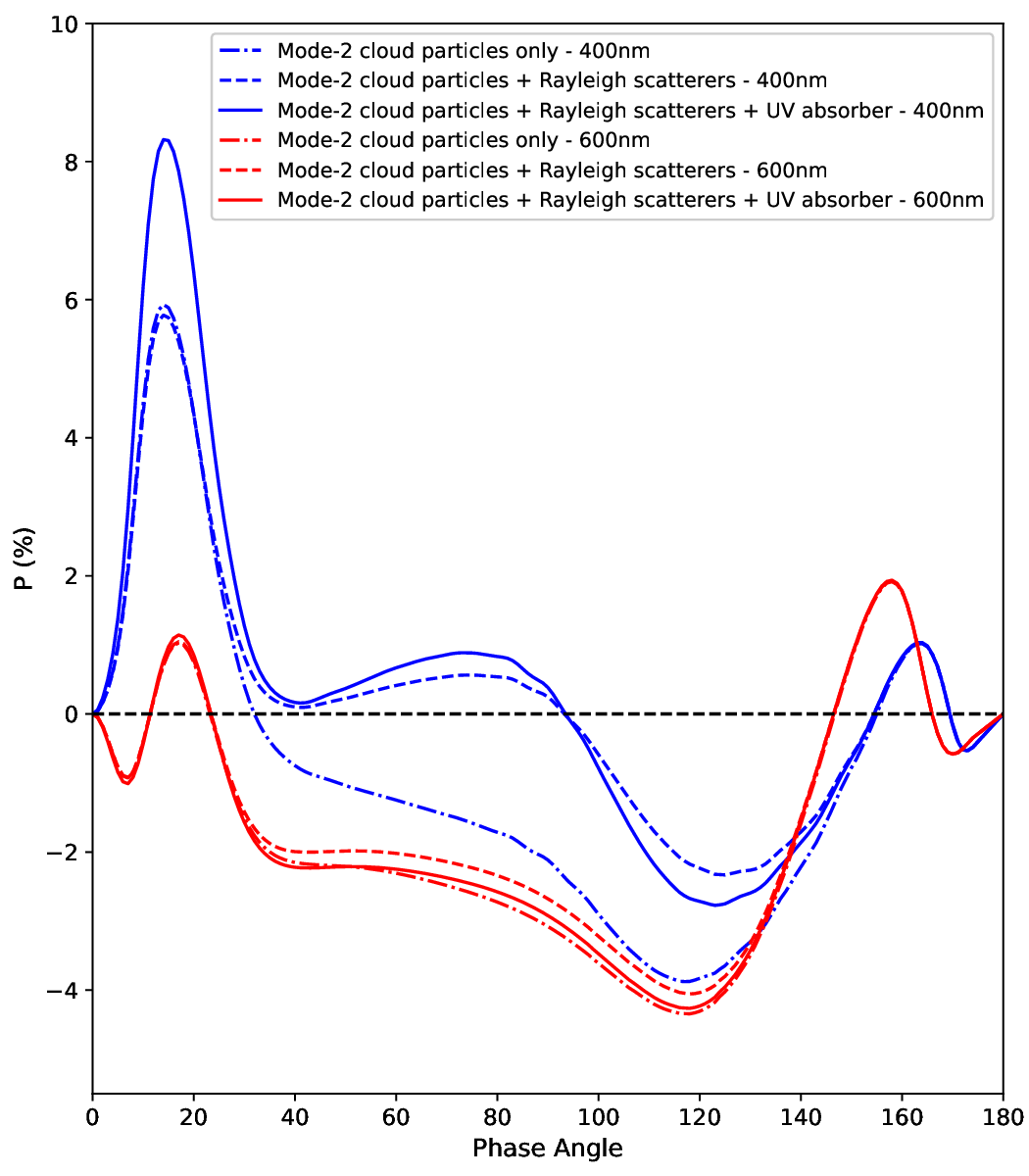}
    \caption{Comparison of the effect of the three components included in the polarization models (see Section \ref{sec:models}). At 400~nm all three components are needed to provide a reasonable match to the observations. At 600~nm the effect of Rayleigh scattering and the UV absorber are relatively small compared with the cloud particles.}
    \label{fig:components}
\end{figure}
  
\subsection{Comparison with past observations and models}

Observations of the polarization of the Venus disk have been reported by Lyot from the 1920s \citep{lyot29}, by Coffeen and Dollfus from the 1950s and 1960s \citep{coffeen65,dollfus70} and from the OCPP instrument on {\it Pioneer Venus} \citep{kawabata80}. The observation sets use a range of different filters and are listed in Table \ref{tab:pastobs}. 

The PICSARR observations are true disk-integrated observations, and the models (both those of HH74 and ours) are disk-integrated models. However, this is not the case for all the observations listed in Table \ref{tab:pastobs}. The observations made with photoelectric aperture polarimeters provide a disk-integrated measurement at most phase angles where the Venus disk is small enough to fit in the aperture. However, at large phase angles, when Venus is in a crescent phase, only the center of the crescent is measured. Observations made with the visual fringe polarimeter measure the center of the disk at all phase angles (defined as the point mid way between the terminator and limb on the equator). 

In Fig. \ref{fig:wavelengths} we present the phase curves in various wavelength regions (UV wavelengths at top left, IR at bottom left, and green/red in the two right panels). As well as the PICSARR observations, we include selected observations from those listed in Table \ref{tab:pastobs} as well as model curves at relevant wavelengths selected from those used in Figs \ref{fig:picsarr_obs} and \ref{fig:mod_wavelengths}. For three of the wavelengths (378 nm, 491 nm and 999 nm) we have included additional models of the polarization at the center of the disk. These are shown as dashed lines, while the corresponding disk-integrated models are solid lines.

Although the overall form of the phase curves is similar in observations and models, Fig. \ref{fig:wavelengths} shows significant differences in detail, both between observations and models, and between the different sets of observations. 

The largest differences seen are for the {\it Pioneer Venus} OCPP data of \cite{kawabata80}. In the UV region the OCPP data (365 nm) is systematically below the models and ground-based data by $\sim$2\%. At 550 nm the OCPP is higher than other observations and models from about 70$^\circ$ < $\alpha$ < 130$^\circ$. At 935 nm the OCPP data does not show the negative excursion reaching $-$4 to $-$5\% at around $\alpha$ = 120$^\circ$ seen in all the other IR data. 

The differences may be in part due to the different view of the planet from the spacecraft compared with that from Earth. The observations were explained with a variable haze of submicron particles (effective radius 0.23 $\mu$m). None of the ground-based data sets show the more positive polarizations seen particularly in the 935 nm OCPP data. \cite{kawabata80} also present separate polarization measurements for polar and equatorial regions that show strongly positive polarization at 935 nm in the polar region.

\begin{table*}[]
     \caption{Past Disk Polarization Observations of Venus}
    \label{tab:pastobs}
   \centering
    \begin{tabular}{llccc}
      Investigator & Instrument & Wavelength & Dates & Disk Integrated?   \\
\hline    Lyot & Visual Fringe Polarimeter & 550nm &   1922-May-22 -- 1924-Jul-01 &    N \\ 
        Dollfus & Photoelectric Aperture Polarimeter & 338nm & 1966-Aug-12 -- 1969-Apr-09 &  Y$^a$ \\
        Dollfus & Photoelectric Aperture Polarimeter & 365nm & 1966-Aug-12 -- 1969-Apr-09 &  Y$^a$ \\
        Dollfus & Photoelectric Aperture Polarimeter & 400nm & 1966-Aug-12 -- 1969-Apr-09 & Y$^a$ \\
        Dollfus & Photoelectric Aperture Polarimeter & 440nm & 1966-Aug-12 -- 1969-Apr-09 & Y$^a$ \\
        Dollfus & Visual Fringe Polarimeter & 440nm & 1959-Oct-14 -- 1966-Jan-17 & N$^b$ \\
        Dollfus & Visual Fringe Polarimeter$^c$ & 527nm & 1950-Sep-24 -- 1969-Apr-09 & N$^b$ \\
        Dollfus & Visual Fringe Polarimeter$^c$ & 593nm & 1950-Oct-04 -- 1969-Apr-09 & N$^b$ \\
        Dollfus & Visual Fringe Polarimeter & 617nm & 1950-Sep-24 -- 1966-Jan-15 & N$^b$ \\
        Dollfus & Photoelectric Aperture Polarimeter & 840nm & 1964-Apr-31 -- 1966-Mar-08 &  Y$^a$ \\
        Dollfus & Photoelectric Aperture Polarimeter & 960nm & 1964-May-05 -- 1966-Mar-08 &  Y$^a$ \\
        Dollfus & Photoelectric Aperture Polarimeter & 1050nm & 1964-May-08 -- 1966-Mar-08 &  Y$^a$ \\
        Coffeen  & Photoelectric Aperture Polarimeter & 340nm & 1961-Jun-29 -- 1969-Sep-03 &  Y$^a$ \\
        Coffeen & Photoelectric Aperture Polarimeter & 364nm & 1959-Apr-23 -- 1969-Aug-21 &  Y$^a$ \\
        Coffeen & Photoelectric Aperture Polarimeter & 445nm & 1959-Oct-15 -- 1969-Jul-05 &  Y$^a$ \\
        Coffeen & Photoelectric Aperture Polarimeter & 520nm & 1961-Oct-16 -- 1969-Jun-25 &  Y$^a$ \\
        Coffeen & Photoelectric Aperture Polarimeter & 550nm & 1959-Apr-24 -- 1969-Apr-18 &  Y$^a$ \\
        Coffeen & Photoelectric Aperture Polarimeter & 655nm & 1965-Aug-01 -- 1969-Jul-04 &  Y$^a$ \\
        Coffeen & Photoelectric Aperture Polarimeter & 685nm & 1959-Oct-15 -- 1962-Sep-11 &  Y$^a$ \\
        Coffeen & Photoelectric Aperture Polarimeter & 740nm & 1964-Jan-25 -- 1968-Jun-30 &  Y$^a$ \\
        Coffeen & Photoelectric Aperture Polarimeter & 875nm & 1964-Jan-25 -- 1969-Jul-04 &  Y$^a$ \\
        Coffeen & Photoelectric Aperture Polarimeter & 990nm & 1959-Apr-23 -- 1969-Jul-04 &  Y$^a$ \\
        Kawabata  & {\it Pioneer Venus} OCPP & 270nm & \multicolumn{1}{l}{Orbits 2-245$^d$} & Y$^e$ \\
        Kawabata  & {\it Pioneer Venus} OCPP & 365nm & \multicolumn{1}{l}{Orbits 2-245$^d$} & Y$^e$ \\
        Kawabata  & {\it Pioneer Venus} OCPP & 550nm & \multicolumn{1}{l}{Orbits 2-245$^d$} & Y$^e$ \\
        Kawabata  & {\it Pioneer Venus} OCPP & 935nm & \multicolumn{1}{l}{Orbits 2-245$^d$} & Y$^e$ \\ 
\hline
    \end{tabular}
\begin{flushleft}
Notes: \\
$^a$ Observations are disk-integrated up to phase angle $\sim$ 140 degrees, but of the center of the crescent at higher angles. \\
$^b$ Observations are of the center of disk (i.e. midway between terminator and limb on equator).\\
$^c$ A small number of high phase angle observations at these wavelengths were made with the photoelectric aperture polarimeter.\\
$^d$ Dates are not given in the paper but the observations cover {\it Pioneer Venus} orbits 2-245 where the orbital period is $\sim$24 hours and orbit was entered on 1978-Dec-4. \\
$^e$ The disk integrated polarization is reconstructed from maps obtained by spacecraft scanning and results in some averaging over a range of phase angles. \\
References: Lyot \citep{lyot29}; Dollfus \citep{dollfus70}; Coffeen \citep{coffeen65,dollfus70}; Kawabata \citep{kawabata80};
\end{flushleft}    
\end{table*}

\subsection{Cloud Particle Size}

One of the main results of HH74 was to determine the properties of the cloud particles. The effects of cloud particle effective radius $r_{\rm eff}$ and effective variance $v_{\rm eff}$ as well as the refractive index were thoroughly investigated in the models presented by HH74. A region of the phase curve that is very sensitive to the particle properties is that including the primary rainbow which causes the polarization peak at phase angles of $\sim$15 to 20 degrees at shorter wavelengths, as well as the negative "glory" feature seen at longer wavelengths. The central phase angle, height and width of these peaks at each wavelength are sensitive to the particle properties.

It is apparent from the data presented in Fig. \ref{fig:picsarr_obs} and \ref{fig:wavelengths} that the new observations fit the modelled rainbow peak quite well. This is particularly apparent from the good fit of the data points for the u$^\prime$ and g$^\prime$ bands to the right edge of the rainbow peak. There is insufficient phase coverage to tie down the left edge of the peak at these wavelengths. However, at the longer wavelengths of the r$^\prime$, i$^\prime$ and z$^\prime$ bands the left edge of smaller, shifted peaks are reasonably well matched.

These observations suggest that the properties of the dominant particle mode have not changed significantly since the 1920s. Whatever is responsible for the long term variability in the phase curves it does not seem to be due to changes in cloud particle size.

\subsection{Comparison with models}

Overall the HH74 models are a good match for the observational data spanning a century. However, as will be seen in the catalogue of discrepancies that follow, there is enough to recommend more sophisticated modelling in order to resolve the many inconsistencies.

\subsubsection{The Ultraviolet Region}

First let us consider the top left panel of Fig. \ref{fig:wavelengths} that compares UV models and observations. As stated above, the available data are a good match for the models in the phase range corresponding to the primary rainbow. Also sensitive to $r_{\rm eff}$ is the so called ``anomalous diffraction peak'' \citep{vandehulst57} produced by interference between diffracted photons and those reflected and transmitted. There are two distinct positions at the right of the phase curve where this peak might occur depending on $r_{\rm eff}$, as seen in Fig. \ref{fig:components}. In Fig. \ref{fig:wavelengths} there is a difference between our new data and the older data presented. The older data are a better match, with the newer data showing a broader peak, or perhaps a double peak. The model peak height -- which is most sensitive to the refractive index -- is, however, a good match to both data sets. Given the older observations are of the centre of the disk at phase angles larger than 140 degrees, it is reasonable to conclude that the polar regions could be responsible for the difference.

We do not wish to over-analyse the marginal discrepancies, but it is worth pointing out that HH74 demonstrate anomalous diffraction is sensitive to particle size distribution and therefore also filter bandwidth, since $r_{\rm eff}$ depends on $\lambda$. The peak becomes washed out if the range is broad. However, we can rule out the filter bandwidth as being responsible for the disagreement, since Fig. \ref{fig:mod_wavelengths} shows a significant shift to lower phase angles occurs only at wavelengths longer than 450 nm -- outside the range of the filter and detector combinations used \citep{coyne67, bailey23}.

The broad second peak depends most strongly on Rayleigh scattering. More Rayleigh scattering particles will increase the peak height and shift its maximum closer to 90 degrees phase. Although the width of the model peak is well matched, the ground-based data consistently shows a peak with greater polarization at lower phase angles than predicted by the models. The OCPP data also displays the Rayleigh scattering peak at lower phase angles, but the polarization is much lower than the model prediction.

\subsubsection{The Visible Region}

The model and observational data are in best agreement in the visual region. The heights of each of the primary rainbow, anomalous diffraction and Rayleigh scattering peaks all agree with the models to within about 0.5 percent. Yet, when one looks closer minor discrepancies begin to emerge; these are most easily seen in the upper-right panel of Fig. \ref{fig:wavelengths}. 

Most noticeably, none of the observations near the 120 degree minimum are as negative as the model predicts, and the slope of the phase curve is straighter than the models predict around 40 degrees -- the ``bridge'' between the primary rainbow and the Rayleigh scattering peak. One could better match the 40 degree feature shape with a larger $v_{\rm eff}$ but fitting the 120 degree minimum requires a smaller $v_{\rm eff}$. A smaller $v_{\rm eff}$ would also significantly increase the height of the primary rainbow peak. The observed polarization in the primary rainbow is already not quite as large as the model predicts, the opposite is true for the anomalous diffraction peak; these two discrepancies are at odds with each other in demanding opposite adjustments to the particle size\footnote{HH74 relied especially on observations of the centre of the crescent to fit their model to the anomalous diffraction peak.}, and point to the need to re-examine the modelling.

\subsubsection{The Infrared Region}

The lower-left panel of Fig. \ref{fig:wavelengths} compares near-infrared models and data. At longer wavelengths the influence of Rayleigh scattering is all but extinguished. The result is good agreement between the models and observation up to phase angles of 90 degrees. Beyond 90 degrees the agreement is poorer than apparent in the UV or Visible region, with the observations producing less negative polarization than the models predict. The PICSARR observations especially show increasing (less negative) polarization at smaller phase angles than predicted by the models. As in the visible region, this points to a need for a smaller $v_{\rm eff}$.

Some of the apparent disagreement can be put down to the change in observing mode, where the observers of the 1960s switched to observing the centre of the disk from 140 degrees onward -- there is better agreement there with disk center model represented by the dashed line. This, of course, does not apply to the PICSARR observations, where for the longest wavelength ($z^\prime$) the shape of the curve is especially poorly reproduced beyond 140 degrees, instead resembling more the 864 nm model.

\subsection{Inter-Synodic Cycle Variability}
\label{sec:seas_var}

There is some evidence for variation between synodic cycles in our data. This is most apparent in Fig. \ref{fig:picsarr_obs} between 40 and 60$^\circ$. We therefore carried out a brief investigation of the scatter of literature data to look for other evidence of this.

\begin{figure}
    \centering
    \includegraphics[width=8.25cm, trim={0 0.1cm 1cm 1cm},clip]{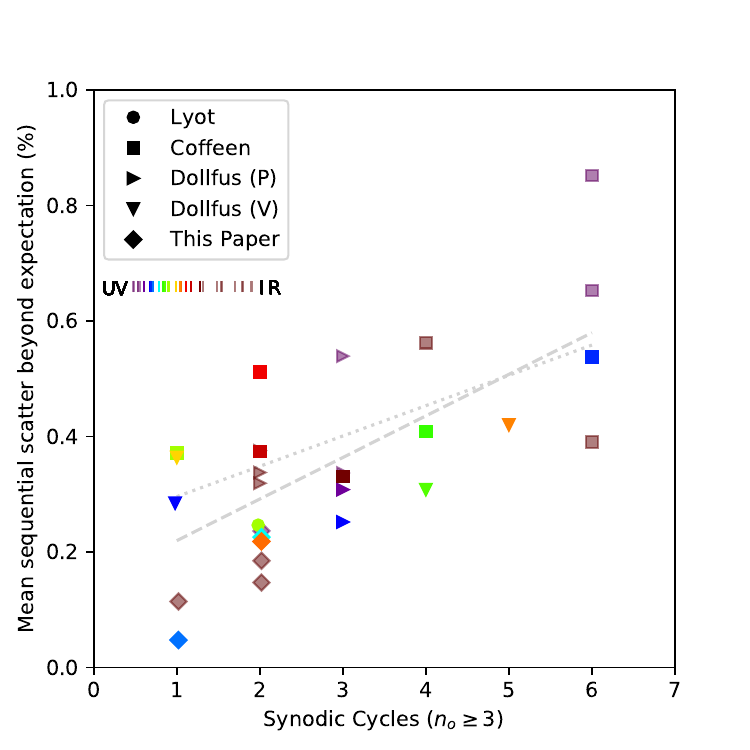}
    \caption{Mean scatter beyond model expectation for historic and current data sets vs number of synodic cycles. Colours indicate wavelength, symbols indicate research lead; V/P indicates visual or photoelectric polarimeter (see key). The dashed and dotted lines represent a linear fit to all and all but the new data in this paper, respectively. The equation of the former is $y=0.072x+0.1477$, with $R^2=0.48$, the equation of the latter is $y=0.0525x+0.2434$, with $R^2=0.38$. More detail is in the text.}
    \label{fig:scatter}
\end{figure}

If there are significant changes between synodic cycles, then we expect datasets covering many cycles to display more scatter in adjacent phase datum. The data presented in Fig. \ref{fig:scatter} represent, for each dataset, the average difference between adjacent phase points minus the expected difference, where the expectation is based on the nearest wavelength model presented in Fig. \ref{fig:mod_wavelengths}; i.e. \begin{equation}\varsigma = \frac{1}{(n_o-1)}\sum_{i=1}^{n} \left [ (p^o_{i+1}-p^o_i) - (p^m_{i+1}-p^m_i) \right ],\end{equation} where $\varsigma$ is the mean sequential scatter beyond expectation; $i$ corresponds to the $i$th element of the observed polarization, $p^o$, in phase order, and $p^m$ is the corresponding model value. For each dataset this is plotted against the number of cycles with at least 3 observations.

A graphical presentation of our analysis (Fig. \ref{fig:scatter}) shows there is greater scatter in data sets including more cycles. A straight line fit gives a rise of 0.072\% per additional cycle, with $R^2=0.48$. However, the trend is likely weaker than it appears, as our new data has much less variation than the other datasets (except \citeauthor{lyot29}'s). Given PICSARR's better reported precision, it is reasonable to expect this is instrument-driven. With the PICSARR data removed, a linear fit gives the rise in scatter as 0.053\% per additional cycle, with $R^2=0.38$.

The polarization of the Venusian atmosphere therefore probably varies noticeably from cycle to cycle. Modern instruments, such as PICSARR, are well equipped to reveal any changes, and should be employed in a long-term program to study the variability, if confirmed.

\subsection{Scatter with Wavelength}

\begin{figure}
    \centering
    \includegraphics[width=8.25cm, trim={0.7cm 0.1cm 1cm 0.7cm},clip]{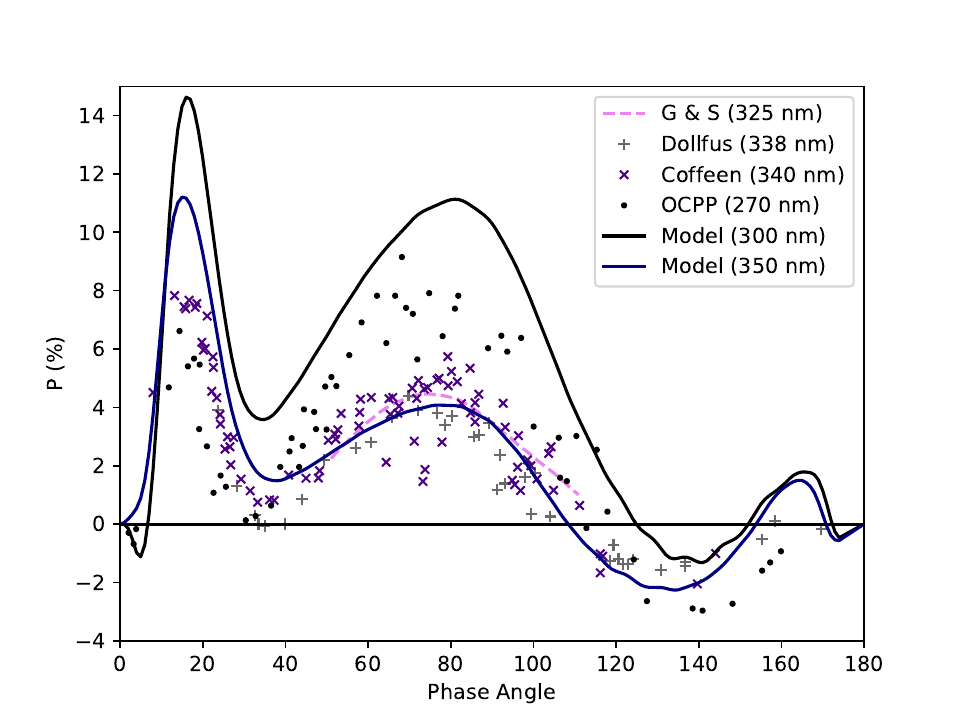}
    \caption{A comparison of data from 270 to 340 nm with models at 300 and 350 nm. Note that the line labelled \mbox{`G \& S} (325 nm)' is data presented as a trend-line only by \citet{gehrels61} for data they say shows significant scatter.}
    \label{fig:UV}
\end{figure}

We performed the same analysis as in Sec. \ref{sec:seas_var}, but looking at the scatter in the data series versus wavelength instead of the number of synodic cycles. No statistically significant trends were revealed by this.

However, it is noteworthy that four of the five datasets with the largest scatter have $\lambda_{\rm eff}\leq445$ nm. The UV data shorter than shown in Fig. \ref{fig:picsarr_obs} is plotted in Fig. \ref{fig:UV}. There is also a trend in our new data -- when considering only the bands observed for more than a full synodic cycle -- for greater scatter at shorter wavelengths. This trend, highlighted in Table \ref{tab:picsarr_scat}, is clear, but small compared to the differences between other data sets. In this context it is interesting to note comments from \citet{gehrels61} based on a partial cycle's observing from 1959 to 1960. They described Venus' polarization as ``remarkably steady from night to night,'' (within the limits of their 0.055\% probable error) but also thought polarization at very short wavelengths may be ``irregularly variable,'' based in part on otherwise unpublished data corresponding to 325 nm. 

\begin{table}
    \centering
    \caption{Scatter in new PICSARR data$^a$.}
    \begin{tabular}{c|c|c}
    \hline
    Filter & Wavelength (nm) & Scatter$^b$ (\%)\\
\hline     
    u$^\prime$  & 378  & 0.273  \\
    g$^\prime$  & 491  & 0.241  \\
    r$^\prime$  & 623  & 0.235  \\
    i$^\prime$  & 759  & 0.185  \\
    z$^\prime_s$/z$^\prime$  & 864/889  & 0.158  \\
    \hline
    \end{tabular}
    \label{tab:picsarr_scat}
\begin{flushleft}
Notes: \\
$^a$ Only filters with more than one synodic cycle of observations are presented. \\
$^b$ Mean scatter beyond expectation between sequential datum ordered by phase. The expected difference between adjacent points is determined by reference to the corresponding models in Table \ref{tab:fil_models}. \\
\end{flushleft}    
\end{table}

Not plotted in Fig. \ref{fig:scatter} (nor used in the Sec. \ref{sec:seas_var} analysis) is the \textit{Pioneer Venus} OCPP data. This is instead reported in Table \ref{tab:ocpp_scat}. Despite corresponding to effectively only one synodic cycle, the scatter is much larger in some filter bands. Whilst the two longer wavelength bands have comparable analogues in Fig. \ref{fig:scatter}, the two UV bands are extreme.

The OCPP instrument acquired disk integrated data, like the photoelectric systems of ground-based observers. However, owing to its orbit, the disk was not the same as the disk seen from Earth, with the polar regions often more prominent. Therefore, greater variability in the polar regions at UV wavelengths could feasibly explain this anomaly. Fig. \ref{fig:UV} demonstrates the scatter, and also poorer agreement of the OCPP measurements with the models compared to other short wavelength data.

\begin{table}
    \centering
    \caption{Scatter in \textit{Venus Pioneer} OCPP data.}
    \begin{tabular}{c|c}
    \hline
    Wavelength (nm) & Scatter$^a$ (\%)\\
\hline     
    270  & 1.146  \\
    365  & 0.587  \\
    550  & 0.292  \\
    935  & 0.289  \\
    \hline
    \end{tabular}
    \label{tab:ocpp_scat}
\begin{flushleft}
Notes: \\
$^a$ Mean scatter beyond expectation between sequential datum ordered by phase. The expected difference between adjacent points is determined by reference to the closest models in Fig. \ref{fig:mod_wavelengths}. \\
\end{flushleft}    
\end{table}

\section{Polarization Images}

\begin{figure*}
    \centering
    \includegraphics[width=17.5cm]{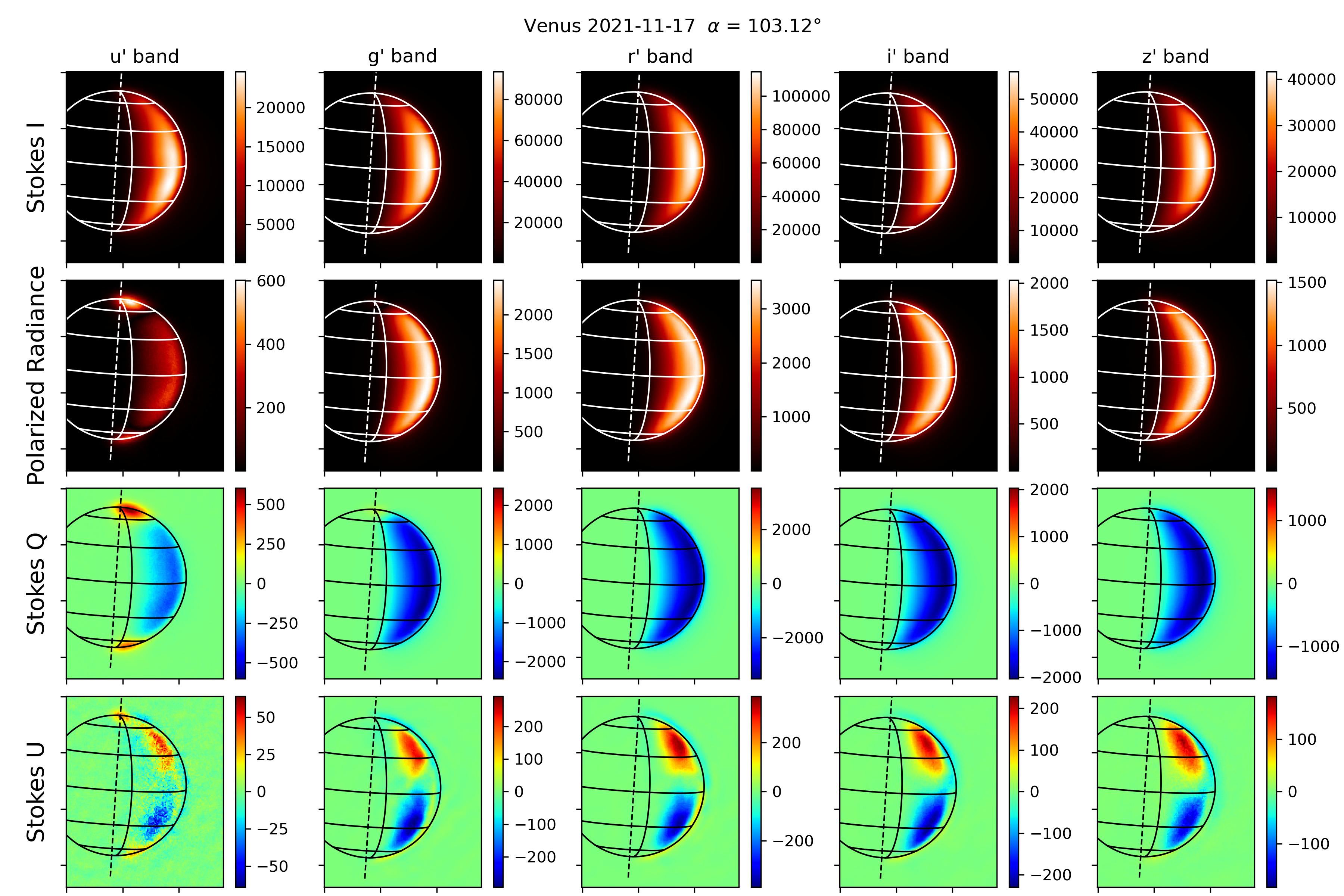}
    \caption{Full set of polarization images on 2021 Nov 17th (Phase angle 103.12$^\circ$) in five filters. The top row shows the Stokes I (radiance) images. The second row shows polarized radiance (= $\sqrt(Q^2 + U^2)$). The lower two rows show the Q Stokes image (polarization perpendicular (+ve) or parallel (-ve) to the scattering plane) and the U Stokes image (polarization at 45$^\circ$ to Q).  The overlay shows the limb of the disk, the planetary rotation axis (dashed line), the terminator, the equator and $\pm$ 30 and 60 degree latitude lines. Fig. \ref{fig:nov17} shows a subset of these images on a larger scale.}
    \label{fig:nov17_2}    
\end{figure*}

\begin{figure*}
    \centering
    \includegraphics[width=17.5cm]{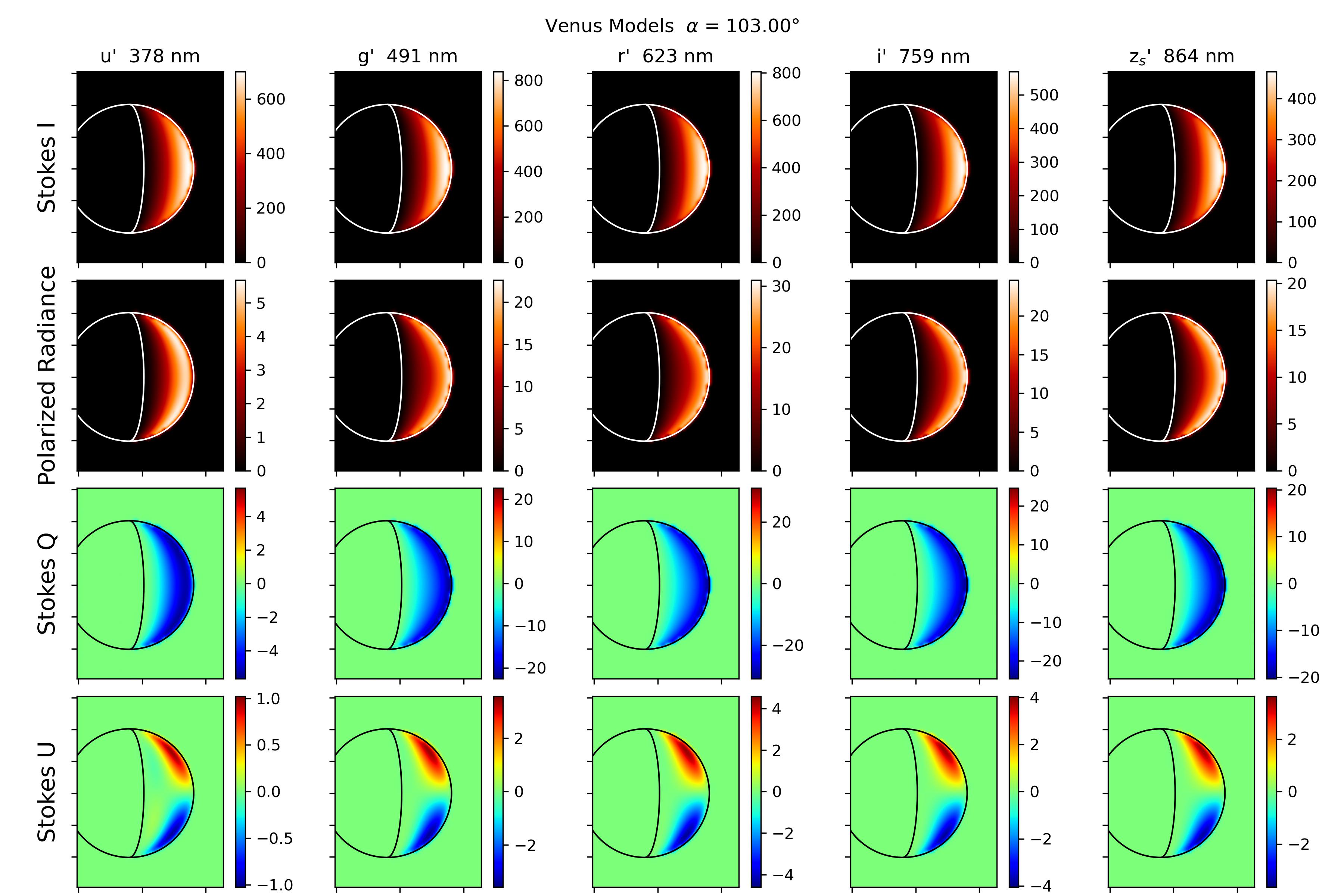}
    \caption{Modelled polarization images of Venus for a phase angle of 103$^\circ$ for comparison with Fig. \ref{fig:nov17_2}. The models used here and in subsequent figures are our reproduction of the HH74 modelling as described in section \ref{sec:repro} and are the same set of models used for the disk integrated data in Fig. \ref{fig:picsarr_obs}. The overlay shows the limb and terminator. Since the model is for a horizontally homogeneous atmosphere the rotation axis and longitude/latitude lines are not meaningful.}
    \label{fig:nov17_model}    
\end{figure*}

\begin{figure*}
    \centering
    \includegraphics[width=17.5cm]{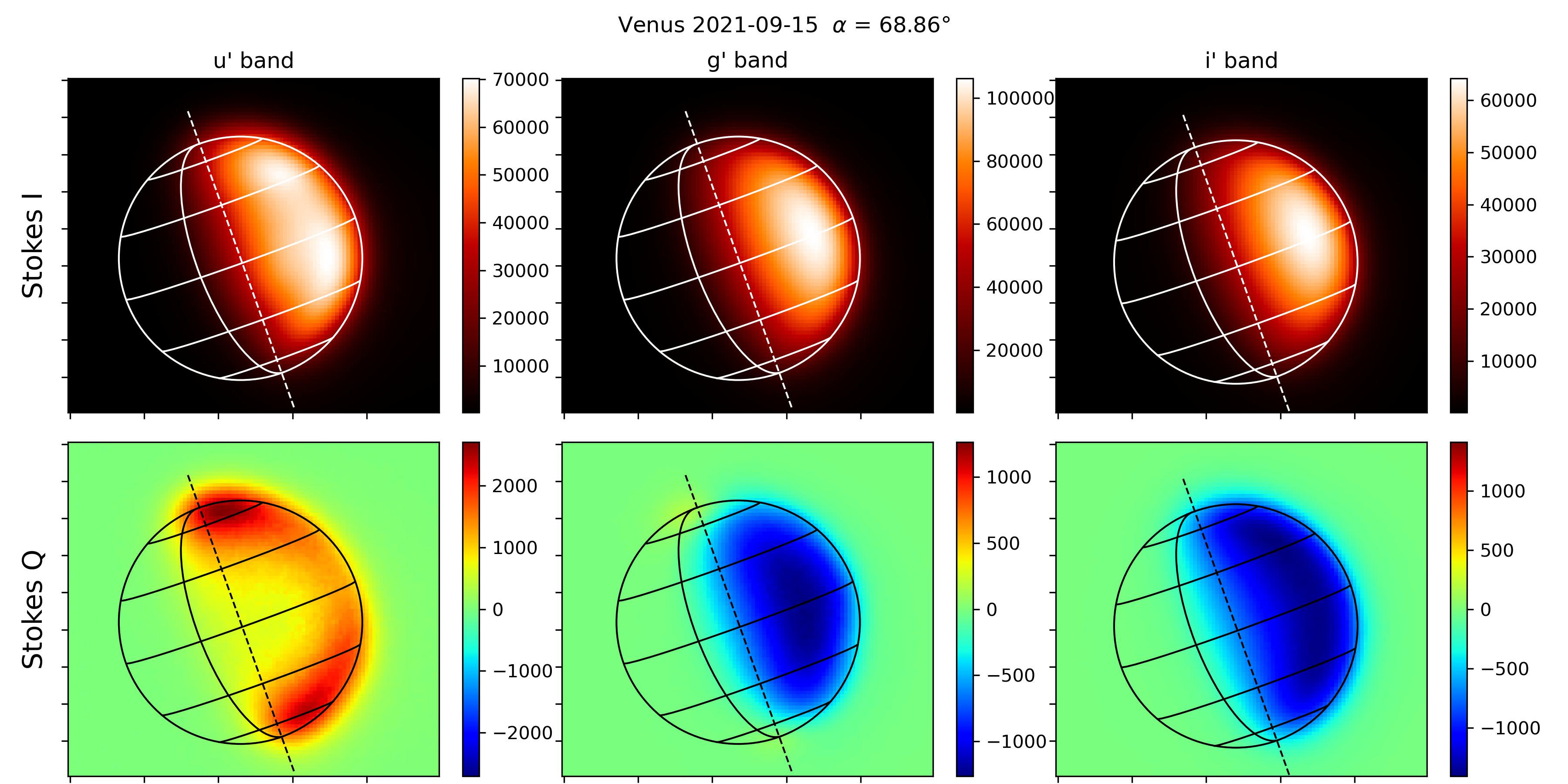}
    \caption{Polarization images on 2021 Sep 15th (Phase angle 68.86$^\circ$) in $u^\prime$, $g^\prime$ and $i^\prime$ bands. The upper panel shows the Stokes I (Radiance) image, and the lower panel is Stokes Q (where positive is perpendicular to the scattering plane). The radiance units are arbitrary, but the same for Stokes Q and Stokes I in each band such that integrating over the image and dividing Stokes Q by Stokes I will give a result equivalent to the polarization shown in Fig. \ref{fig:picsarr_obs}. Celestial north is at top. The overlay shows the limb of the disk, the planetary rotation axis (dashed line), the terminator, the equator and +/$-$ 30 and 60 degree latitude lines.}
    \label{fig:sep15}    
\end{figure*}

\begin{figure*}
    \centering
    \includegraphics[width=17.5cm]{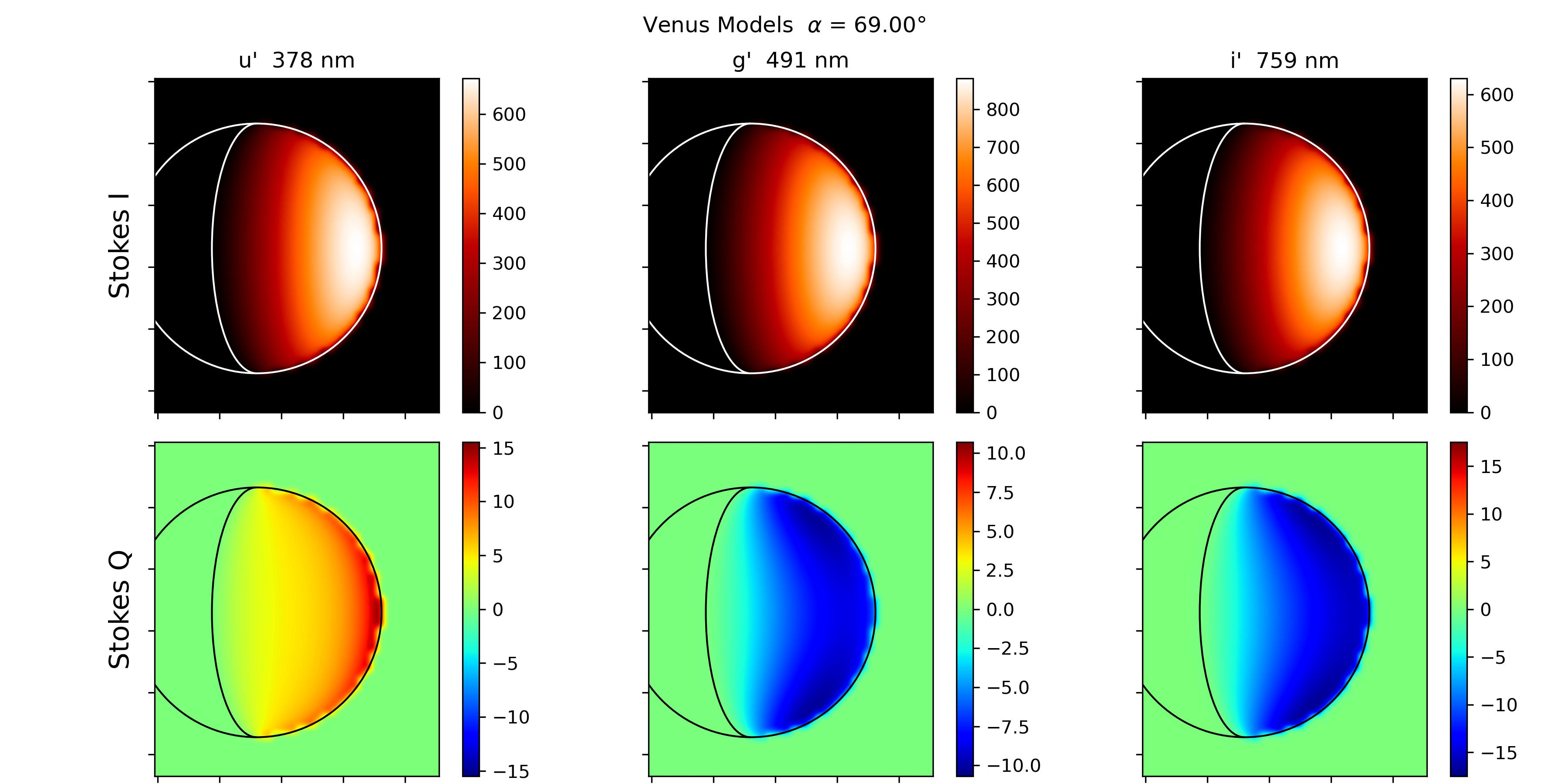}
    \caption{Modelled polarization images of Venus for a phase angle of 69$^\circ$ for comparison with Fig. \ref{fig:sep15}. }
    \label{fig:mod_ph69}    
\end{figure*}

\begin{figure*}
    \centering
    \includegraphics[width=17.5cm]{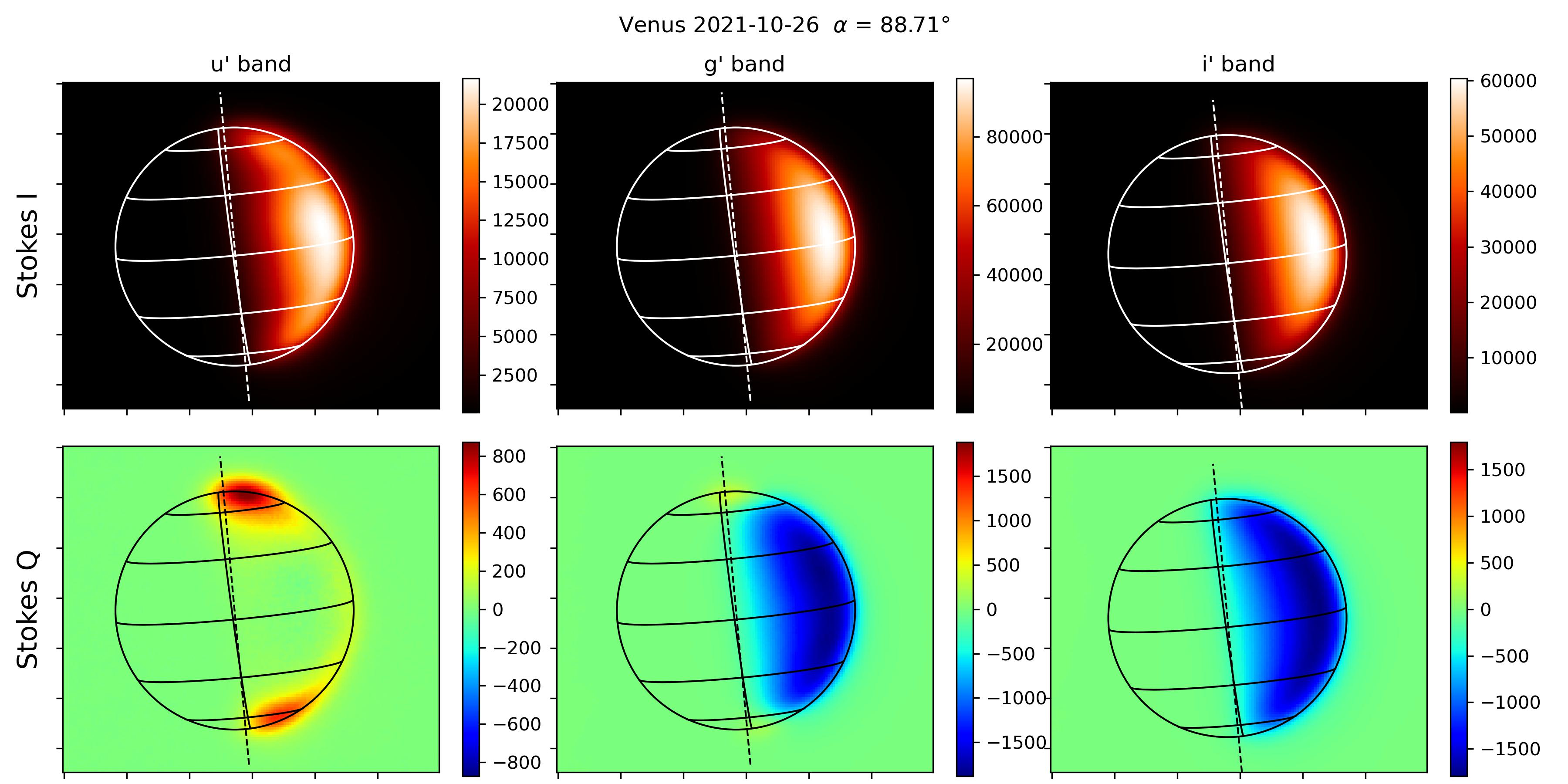}
    \caption{As Fig. \ref{fig:sep15} for 2021 Oct 26th (Phase angle 88.71$^\circ$).}
    \label{fig:oct26}    
\end{figure*}

\begin{figure*}
    \centering
    \includegraphics[width=17.5cm]{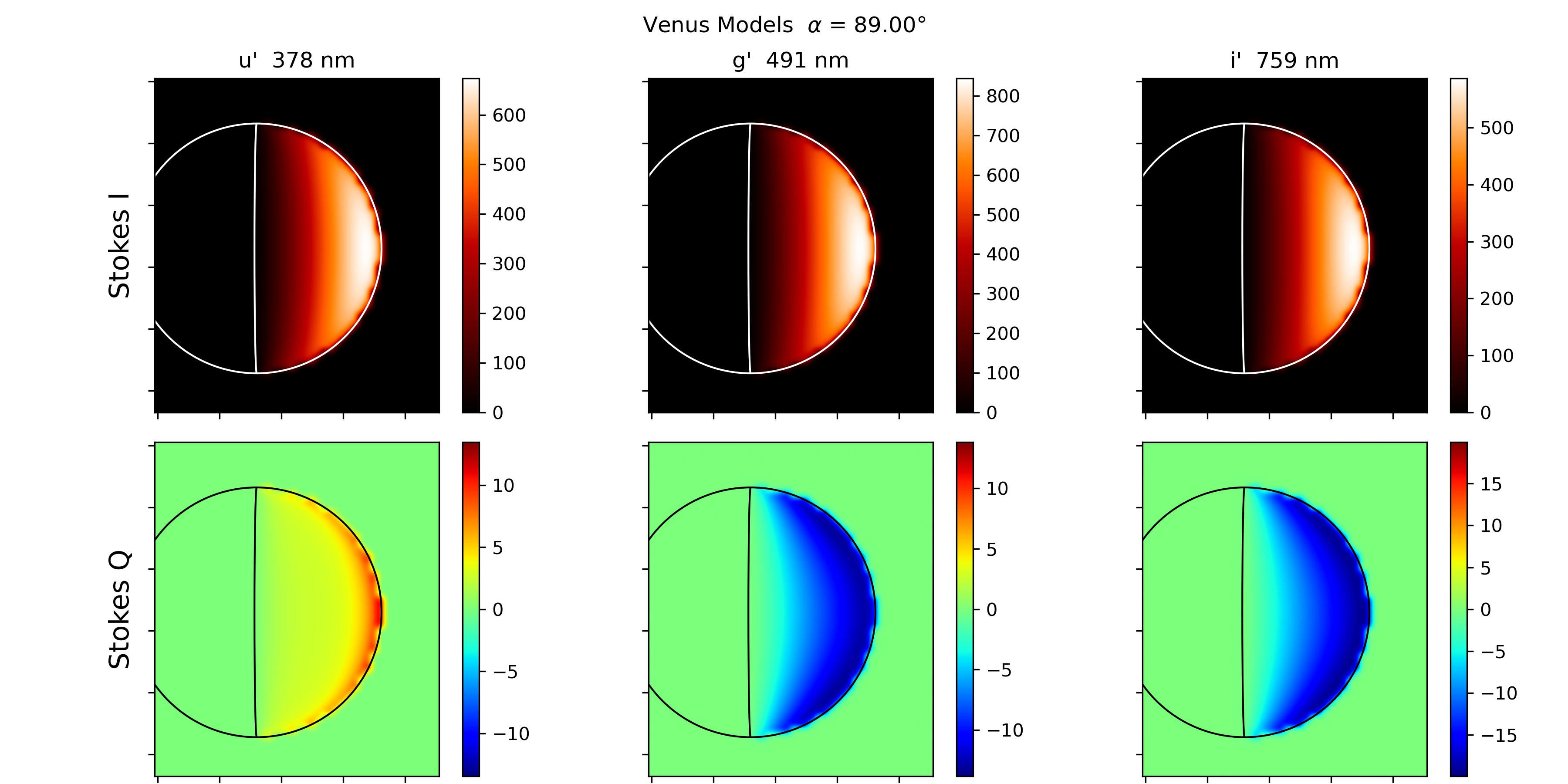}
    \caption{Modelled polarization images of Venus for a phase angle of 89$^\circ$ for comparison with Fig. \ref{fig:oct26}.}
    \label{fig:mod_ph89}    
\end{figure*}

\begin{figure*}
    \centering
    \includegraphics[width=17.5cm]{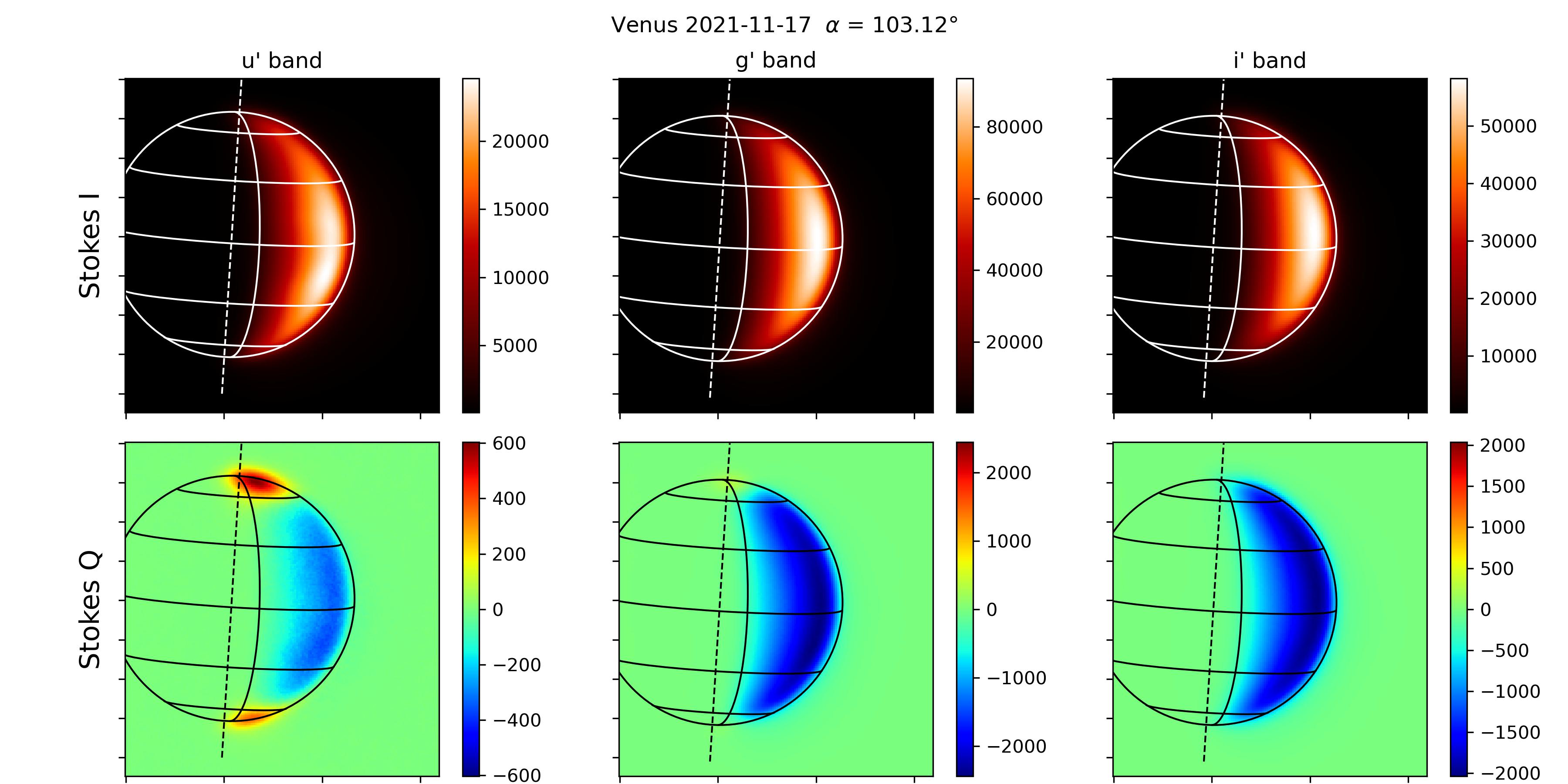}
    \caption{As Fig. \ref{fig:sep15} for 2021 Nov 17th (Phase angle 103.12$^\circ$). The $u^\prime$ band disk-integrated polarization crosses from positive to negative at about this phase in both observations and models. However, the image shows that the polarization is still strongly positive at polar latitudes while becoming negative at lower latitudes.}
    \label{fig:nov17}    
\end{figure*}

\begin{figure*}
    \centering
    \includegraphics[width=17.5cm]{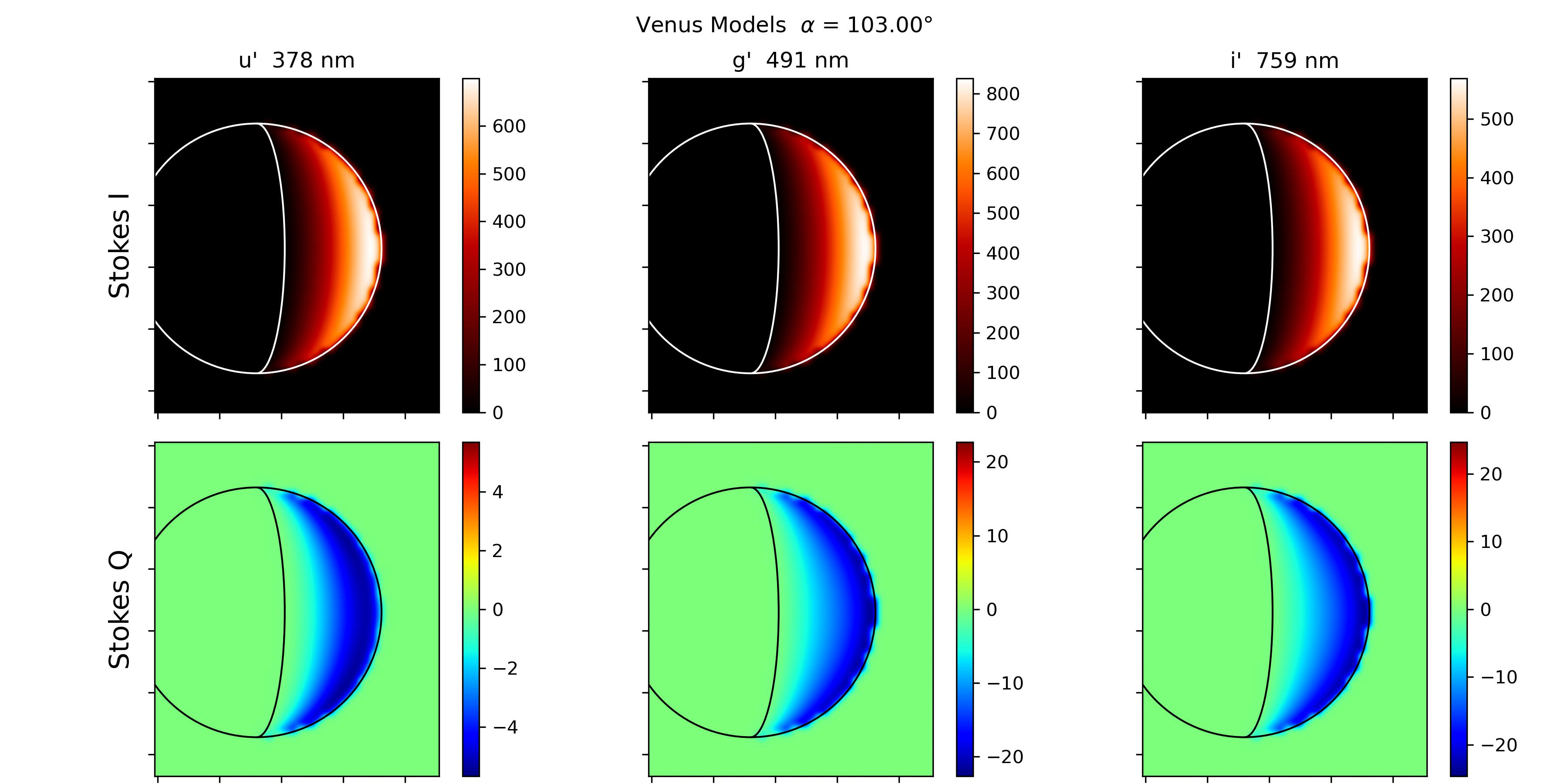}
    \caption{Modelled polarization images of Venus for a phase angle of 103$^\circ$ for comparison with Fig. \ref{fig:nov17}.}
    \label{fig:mod_ph103}    
\end{figure*}

\begin{figure*}
    \centering
    \includegraphics[width=17.5cm]{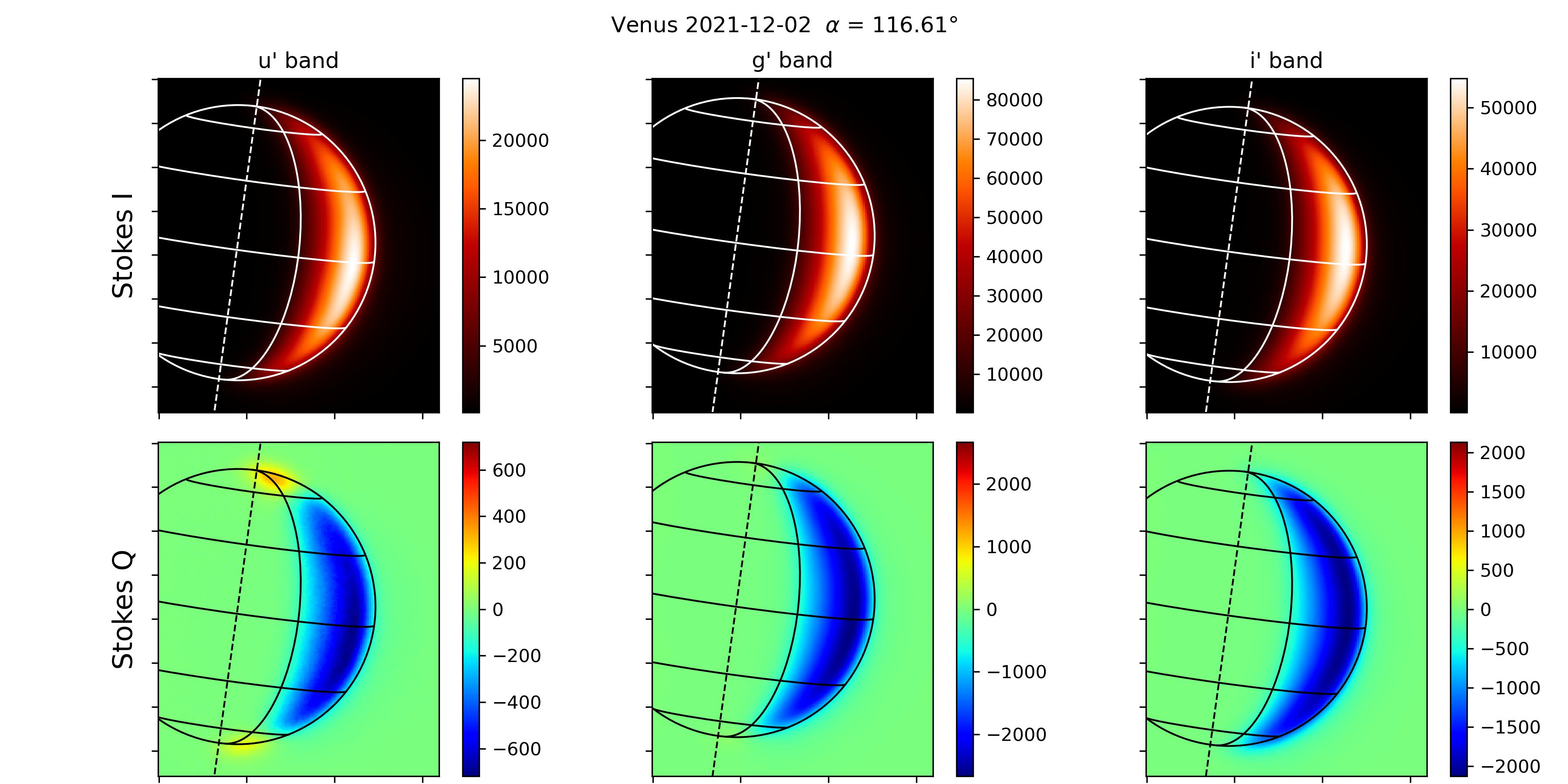}
    \caption{As Fig. \ref{fig:sep15} for 2021 Dec 2nd (Phase angle 116.61$^\circ$).}
    \label{fig:dec02}    
\end{figure*}

\begin{figure*}
    \centering
    \includegraphics[width=17.5cm]{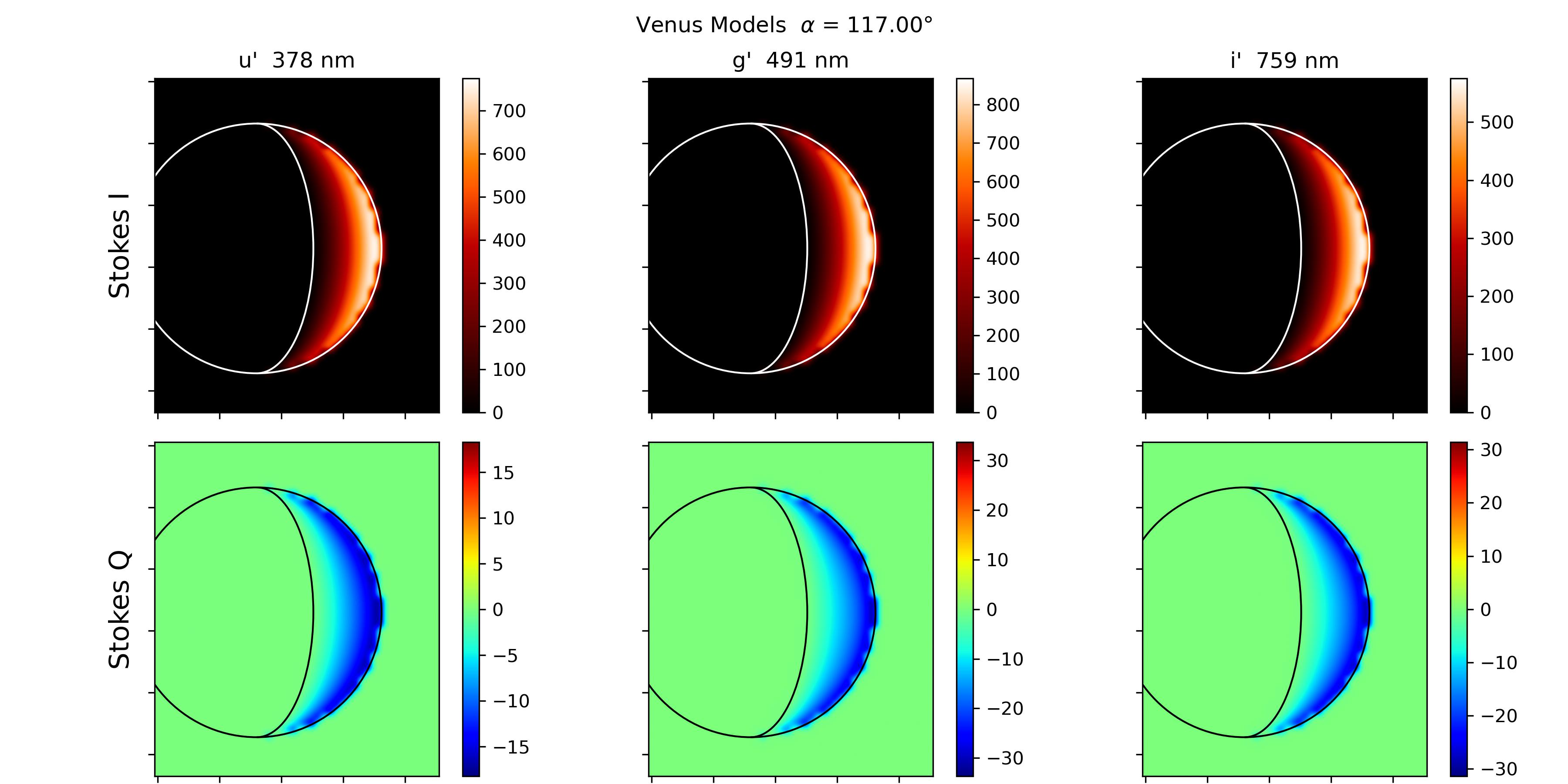}
    \caption{Modelled polarization images of Venus for a phase angle of 117$^\circ$ for comparison with Fig. \ref{fig:dec02}.}
    \label{fig:mod_ph1117}    
\end{figure*}

Past observations of the regional variation of polarization across the disk of Venus have been made by mapping the planet using many individual observations made with a visual fringe polarimeter or a photoelectric aperture polarimeter \citep{dollfus79}. Temporary patches of anomalous polarization were observed and interpreted as local changes in particle size or cloud top altitude.

PICSARR data can be used to derive images of the polarization variation across the planet. As we are using small telescopes at low-altitude sites, and since Venus often has to be observed at low elevations where seeing is poor, the imaging data are often of limited quality. The best observations for this purpose are those obtained when Venus is near elongation (phase angles around 90 degrees) so that Venus can be observed at a relatively high elevation in a dark sky. Favourable observing conditions occurred from September to December 2021. Further observations were obtained over the same range of phase angles in 2023 and show essentially the same behavior as seen in 2021.

Fig. \ref{fig:nov17_2} shows an example of the full set of images obtainable from observations in the five filters. The images are in the form of the Stokes parameters (I, Q and U). As for the disk-integrated case, polarization is referenced to the scattering plane --- the plane containing the Sun, planet and observer. The Q Stokes parameter is positive for polarization perpendicular to the scattering plane and negative for polarization parallel to the scattering plane. The U Stokes parameter is a measure of polarization at $\pm45$ degrees to the scattering plane. 

The images are plotted with celestial north at top and east at left. The orientation of the planet according to the JPL Horizons Ephemeris System \citep{giorgini15} is indicated by an overlay that shows the limb of the planet, the rotation axis, the terminator, the equator and the $\pm30$ and 60 degree latitude lines. For these observations there is a small tilt of the rotation axis towards the observer so that the north pole is visible and the south pole is hidden. There is also a small tilt of the rotation axis towards the Sun so that the north pole is illuminated and the south pole is in shadow.

We can also use the models described in section \ref{sec:repro} to predict the polarization distribution across the planetary disk for comparison with the imaging observations. The models corresponding to the observations in Fig. \ref{fig:nov17_2} are shown in Fig. \ref{fig:nov17_model}. The modelled geometry does not exactly match the observations since it does not include the tilt of the illumination direction relative to the celestial coordinate system.

\subsection {The U Stokes polarization}

The polarization images such as those in Fig. \ref{fig:nov17_2} reveal a new feature of the polarization of Venus that has not previously been described --- the presence of polarization in the U Stokes parameter.

As described by HH74 \citep{hh74}, the polarization of Venus is dominated by single scattering. Sunlight most often reaches us after scattering off a single particle in the atmosphere. In this case the symmetry of the situation requires that the polarization can only be perpendicular or parallel to the scattering plane and hence only in the Q Stokes parameter as we have defined it.

There is no such limitation on light that has undergone multiple scattering, which can contribute to polarization in both the Q and U Stokes parameters \citep{vanderhulst80}. Any polarization seen in the U Stokes parameter must be the result of multiple scattering. 

Multiple scattering polarization in planetary atmospheres has been observed and described most commonly for the outer planets observed near zero phase angle (i.e. near opposition) at which point the single scattering polarization is usually small and the multiple scattering effects can be dominant. At zero phase angle the polarization takes the form of polarization increasing from the centre to the limb and perpendicular to the limb. This limb polarization has been observed in Uranus and Neptune \citep{schmid06,joos07}, Jupiter and Saturn \citep{schmid11} and Titan \citep{bazzon14}. For a planet with a horizontally homogeneous atmosphere the multiple scattering polarization at zero phase angle produces a ring of high polarization around the limb of the planet and a distinctive ``four-lobed'' or ``butterfly'' pattern of alternating positive and negative signs in the Q and U Stokes parameters \citep{schmid06, bazzon14}. The disk integrated polarization of such a pattern is zero.

For our Venus observations at higher phase angles multiple scattering is still present and results in polarization in the U Stokes parameter that can be seen in the observations (Fig. \ref{fig:nov17_2}) and models (Fig. \ref{fig:nov17_model}). The polarization is about an order of magnitude lower in Stokes U than in Stokes Q. The modelled Stokes U images have positive and negative features either side of the scattering plane such that the U contribution will average to zero and not contribute to the disk-integrated polarization.

\begin{figure}
    \centering
    \includegraphics[width=8.25cm]{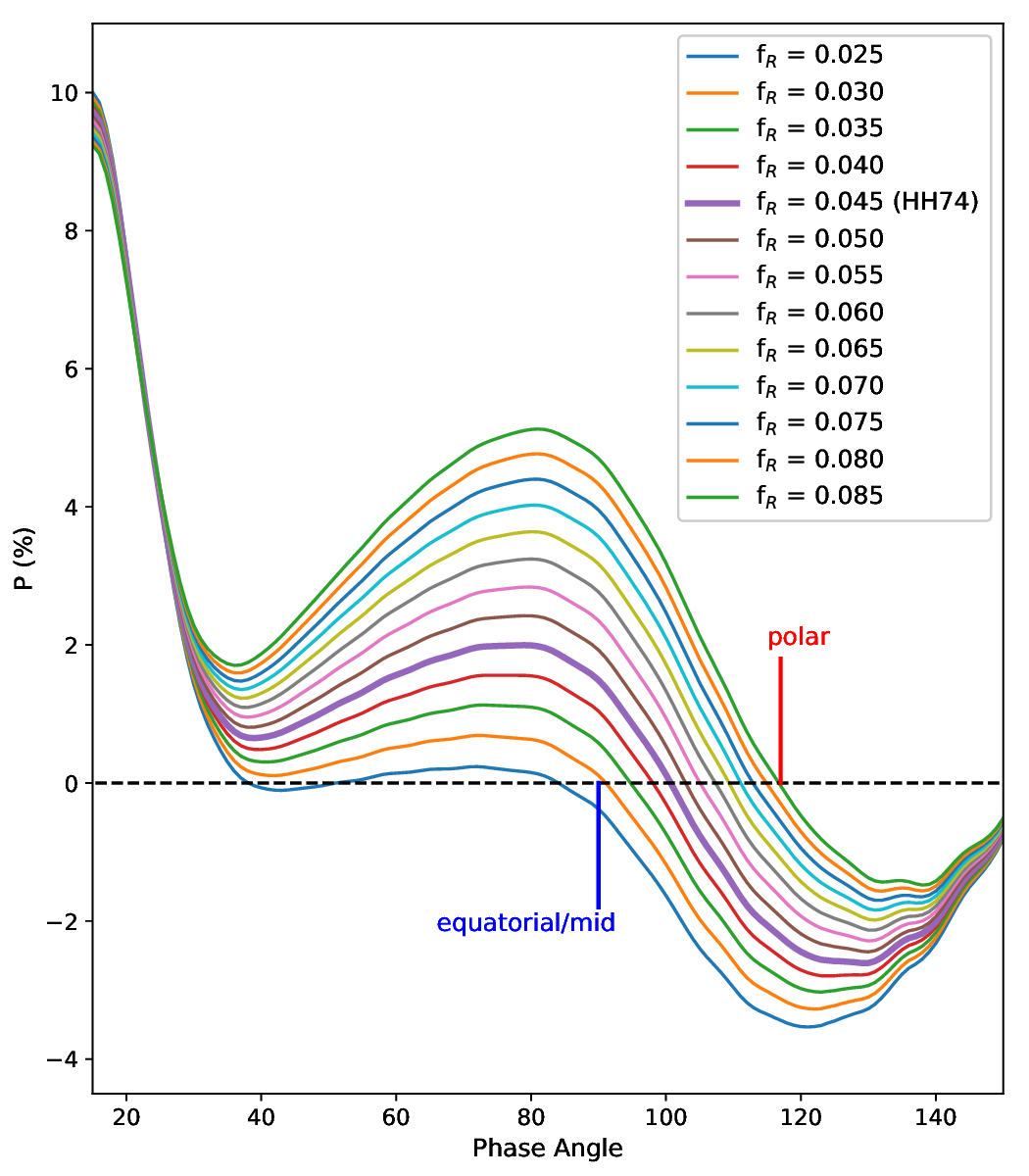}
    \caption{Models of the 378nm (u$^\prime$ band) polarization for different values of the Rayleigh scattering fraction (f$_R$). The standard HH74 value is f$_R$ = 0.045. Markers show the estimated zero crossing phase angles from the imaging data (Figures \ref{fig:sep15}, \ref{fig:oct26}, \ref{fig:nov17}, \ref{fig:dec02}) for the equatorial and mid-latitudes (at f$_R$$\sim$0.030) and the polar latitudes (at f$_R$$\sim$0.085).} 
    \label{fig:venus_fr}
\end{figure}

\subsection{Polarization features in the polar regions}

In figures \ref{fig:sep15} to \ref{fig:mod_ph1117} we show observations and corresponding models at four phase angles from 69$^\circ$ to 117$^\circ$. The observations shown here are obtained in the u$^\prime$, g$^\prime$ and i$^\prime$ filters. As can be seen in Fig. \ref{fig:nov17_2} and \ref{fig:nov17_model} there is little difference between the $r^\prime$, i$^\prime$ and z$^\prime$ filters in either the observations or models, so the i$^\prime$ filter is representative of all three bands.

The observed and modelled distribution of Stokes I are quite similar. They are not identical, however, with the observations generally showing more of a fall-off in radiance near the limb and terminator of the planet. This is likely to be a combination of factors. The models are plane-parallel models and therefore become less accurate when either the observed or solar viewing angle approaches 90 degrees, as is the case at the limb and terminator. They are also simplified in their representation of the atmosphere with a single optically-thick layer. Also the observations are blurred by atmospheric seeing and other instrumental imaging issues. Typically, the seeing at our low-altitude observing site is $\sim$2 arc-seconds at best, and the image diameters range from 17 to 39 arc-seconds. In the case of the $u^\prime$ band observations there can be additional structure in the Stokes I image due to UV absorbing clouds as is apparent in Fig. \ref{fig:sep15} and \ref{fig:oct26}.

In the Stokes Q images there is also reasonable agreement between the observations and model in the i$^\prime$ band, subject to the same issues mentioned for Stokes I. In particular all the imaging observations at $r^\prime$, $i^\prime$ and $z^\prime$ show no evidence for the much stronger polarization in the polar regions that was reported for the OCPP data \citep{kawabata80}. The strong haze of submicron particles proposed by \cite{kawabata80} does not seem to be present during our observations. However the SPICAV-IR data \citep{rossi15} obtained from 2006-2010 does show high IR polarization at high latitudes similar to that reported by \cite{kawabata80}.

However, in the u$^\prime$ band the Stokes Q image shows features in the polar regions that are quite different from the models. There are regions of strongly positive polarization near the north and south poles where the rest of the planet is much less positive (Fig. \ref{fig:sep15} and \ref{fig:oct26}) or shows negative polarization (figure \ref{fig:nov17} and \ref{fig:dec02}). Although much weaker than in u$^\prime$ a similar effect is apparent in g$^\prime$. In Figure \ref{fig:oct26} the blue region of negative polarization stops at the 60$^\circ$ north line, rather than extending nearly to the north pole as it does at i$^\prime$. There is a slightly positive region very close to the pole.

From the disk integrated models (Fig. \ref{fig:picsarr_obs}) it can be seen that the modeled polarization in u$^\prime$ band (378nm) crosses from positive to negative at a phase angle of 100 degrees, and the imaging models (Fig. \ref{fig:mod_ph89} and \ref{fig:mod_ph103}) indicate that the whole of the disk switches from positive to negative at about the same phase angle. The observations show that the local behavior is quite different to this. In the equatorial and mid-latitudes the negative-to-positive transition occurs at about phase angle 90 degrees, whereas in the polar regions the transition is not quite complete in Fig. \ref{fig:dec02} and occurs at a phase angle of about 117 degrees.

\subsection{Interpretation of the polar polarization behavior}

In the context of the HH74 models the transition from positive to negative polarization in u$^\prime$ band is primarily determined by f$_R$, the fraction of Rayleigh scatterers included in the model as described in Section \ref{sec:models}. The adoption of the value f$_R$ = 0.045 by HH74 \citep{hh74} was largely based on matching this transition in the ultraviolet data. Because Rayleigh scattering is strongly wavelength dependent, this parameter has much less effect at longer wavelengths (see Fig. \ref{fig:components}).

The observed polar polarization behavior can then be understood if there is a larger Rayleigh scattering fraction (f$_R$) in the polar regions than in the equatorial and mid latitudes, rather than a single value over the whole planet. Fig. \ref{fig:venus_fr} shows a set of models for the u$^\prime$ (378nm) polarization phase curve which use the standard HH74 parameters except for a changing value of f$_R$. When these results are compared with the imaging data it can be seen that the positive to negative crossing for the equatorial and mid latitudes at about 90 degrees corresponds to f$_R$ $\sim$ 0.030, and that for the polar latitudes at 117 degrees corresponds to f$_R$ $\sim$ 0.085.

HH74 \citep{hh74} interpret the f$_R$ parameter as Rayleigh scattering due to atmospheric gas, with the standard value of 0.045 corresponding to a cloud-top pressure of 50 mbar. In the same way our results can be interpreted as requiring a cloud top pressure of $\sim$33 mbar in the equatorial and mid latitudes, and $\sim$94 mbar in the polar regions.   Using the relationships between pressure and altitude in the Venus International Reference Atmosphere \citep{seiff85} this corresponds to a $\sim$6 km difference in cloud top altitude from 70 km at equatorial/mid latitudes to 64 km in the polar regions.

Such an interpretation is in reasonable agreement with a number of other studies. Akatsuki IR2 results \citep{sato20} give a cloud top altitude of 70.5 km at the equator, a rapid drop at latitudes of 50-60 degrees, and a cloud-top at 61 km in latitudes 70-75 degrees. From SPICAV/Venus Express observations \citep{federova16} the average altitude is given as 70.2 $\pm$ 0.8 km at equatorial latitudes decreasing to 62-68 km at high latitudes. Observations with Venus Express VeRa and VIRTIS instruments \citep{lee12} show the cloud top at 67.2 $\pm$ 1.9 km in low latitudes and at 62.8 $\pm$ 4.1 km at the pole.

\section{Conclusions}

We have obtained new observations of the polarization of Venus over three years from August 2021 to October 2024. The observations were obtained with small telescopes (20cm and 35cm apertures) and PICSARR polarimeters
\citep{bailey23} that provide high precision and the ability to obtain disk-integrated and imaging data. We compare the results with previous observations that go back to the work of \cite{lyot29}.

We have reproduced the HH74 models \citep{hh74} of the polarization of Venus using modern radiative transfer codes \citep{bailey18}. We show that the new models are in good agreement with the originals, and we are able to calculate results for our filter wavelengths that are different from those used in the original work. We are also able to calculate models of the polarization distribution across the disk, to compare with imaging data.

The new disk-integrated observations are broadly in agreement with past observations and with the models. The observations agree with past determinations of the size distribution of the predominant particle mode. They agree with past observations in showing evidence for variability of the phase curves on long and short timescales. There are also differences between observations and models in the detailed structure of the phase curves. These differences cannot be easily explained by simple changes to the main parameters of the HH74 type models, and indicate the need for more sophisticated models.

Imaging observations show a distribution of polarization across the disk of Venus that agrees with the model predictions in the redder filters (r$^\prime$, i$^\prime$, z$^\prime$). However, in the u$^\prime$ filters we see very different polarization in the polar regions (within about 30 degrees of the poles) compared with that at equatorial and mid-latitudes. The more positive polarization in the polar regions can be understood if there is a higher contribution of Rayleigh scatterers in these regions and this could be explained by $\sim$6 km lower cloud top altitude (64 km at the poles compared with 70 km at equatorial and mid latitudes). While this is in agreement with past results from spacecraft observations, the ability to measure such effects with small ground-based telescopes opens up the possibility of more frequent monitoring and mapping of the cloud-top altitude. 

We do not see significant latitudinal variations in polarization at red and IR wavelengths. This is in contrast to the large positive polarization at high latitudes reported in the OCPP data of \cite{kawabata80} and the SPICAV-IR data of \cite{rossi15}. This strongly suggests that in our three years of coverage with PICSARR we have not seen the full range of variability that can occur in the Venus atmosphere. Continued observations with PICSARR or similar instruments are desirable. It would be particularly useful to obtain further imaging polarimetry with somewhat larger telescopes on high-altitude good-seeing sites that could obtain higher spatial resolution over a wider range of phase angles.

The models used in this paper are models of a horizontally homogeneous atmosphere - i.e. one that has the same structure at every latitude and longitude on the planet. Our observations, particularly the u$^\prime$ band polarization images, show that the actual Venus atmosphere is not horizontally homogeneous. In the future we plan to investigate the polarization properties of Venus using more detailed models that take into account the latitudinal differences in the atmosphere and use a multi-layer multi-mode cloud model in better agreement with current understanding of the Venus clouds.



\printcredits

\section*{Declaration of competing interests}

The authors declare that they have no known competing financial interests or personal relationships that could have appeared to influence the work reported in this paper.

\section*{Acknowledgments}

KB acknowledges the support of NASA Habitable Worlds grants No. 80NSSC24K0074 \& No. 80NSSC20K1529. DVC and IB thank the Friends of MIRA.

\section*{Data Availability}

The table of disk integrated observations is available on the Mendeley Data repository \citep{Bailey25}. Other data can be made available on request.

\bibliographystyle{cas-model2-names}

\bibliography{venus_pol}





\end{document}